\documentclass{emulateapj}
\usepackage{amsmath, amssymb}
\usepackage{color}
\usepackage[normalem]{ulem}

\slugcomment{Not to appear in Nonlearned J., 45.}

\shorttitle{Stochastic Acceleration in Pulsar Wind Nebulae}
\shortauthors{S. J. Tanaka, \& K. Asano}

\begin{document}


\title{On the Radio Emitting Particles of the Crab Nebula: Stochastic Acceleration Model}


\author{Shuta J. Tanaka\altaffilmark{1}, \& Katsuaki Asano\altaffilmark{2}}
\altaffiltext{1}{
	Department of Physics, Faculty of Science and Engineering, Konan University, 8-9-1 Okamoto, Kobe, Hyogo 658-8501, Japan
}
\altaffiltext{2}{
	Institute for Cosmic Ray Research, The University of Tokyo, 5-1-5 Kashiwa-no-ha, Kashiwa City, Chiba, 277-8582, Japan
}
\email{sjtanaka@center.konan-u.ac.jp}
%
%




\begin{abstract}
The broadband emission of Pulsar Wind Nebulae (PWNe) is well described by non-thermal emissions from accelerated electrons and positrons.
However, the standard shock acceleration model of PWNe does not account for the hard spectrum in radio wavelengths.
The origin of the radio-emitting particles is also important to determine the pair production efficiency in the pulsar magnetosphere.
Here, we propose a possible resolution for the particle energy distribution in PWNe; the radio-emitting particles are not accelerated at the pulsar wind termination shock but are stochastically accelerated by turbulence inside PWNe.
We upgrade our past one-zone spectral evolution model including the energy diffusion, i.e., the stochastic acceleration, and apply to the Crab Nebula.
A fairly simple form of the energy diffusion coefficient is assumed for this demonstrative study.
For a particle injection to the stochastic acceleration process, we consider the continuous injection from the supernova ejecta or the impulsive injection associated with supernova explosion.
The observed broadband spectrum and the decay of the radio flux are reproduced by tuning the amount of the particle injected to the stochastic acceleration process.
The acceleration time-scale and the duration of the acceleration are required to be a few decades and a few hundred years, respectively.
Our results imply that some unveiled mechanisms, such as back reaction to the turbulence, are required to make the energies of stochastically and shock accelerated particles comparable.
\end{abstract}





\section{INTRODUCTION}\label{sec:intro}

Pulsar wind nebulae (PWNe) are a kind of supernova (SN) remnants having characteristics of a `filled-center morphology' and a `flat radio spectrum' \citep[][]{Weiler&Panagia78}.
The filled-center morphology results from plasmas supplied by a central pulsar of a PWN.
The relativistic outflow from a pulsar, called pulsar wind, interacts with external SN ejecta and forms a termination shock \citep[c.f.,][]{Rees&Gunn74, Kennel&Coroniti84a}.
We observe non-thermal radiation from the particles accelerated at the termination shock, while the standard magnetohydrodynamic model of PWNe does not account for the characteristic flat radio spectrum \citep[see \S Vd of][]{Kennel&Coroniti84b}.

The flat radio spectra raise some problems in the origin of the radio emission.
One is that the spectrum of the `radio-emitting particles' is too hard to be explained by the diffusive shock acceleration \citep[c.f.,][]{Blandford&Eichler87}.
The typical spectral index of the required electron distribution is 1--1.6, which is smaller than the value $2$ in the standard shock acceleration theory,
but softer than the Maxwellian distribution.
To generate such a hard spectrum, another kind of acceleration mechanism such as magnetic reconnection \citep{Sironi&Spitkovsky11} may be required.
On the other hand, the X-ray spectrum of PWNe is soft enough to be explained by the shock accelerated particles \citep[][]{Kennel&Coroniti84b}.
The other problem is that the radio-emitting particles dominate in the particle number inside PWNe and are too much compared to the estimated number
of particles supplied by the pulsar wind.
The number flux of the pulsar wind is estimated from the pair cascade process in the pulsar magnetosphere \citep[e.g.,][]{Timokhin&Harding15} and any theoretical predictions of the pair amount are at least one to two orders of magnitudes less than the number estimated from the radio observation \citep[c.f.,][]{Arons12, Tanaka&Takahara13a}.
The formation of the characteristic flat radio spectrum is an indispensable issue to understand the particle acceleration processes and the physics around pulsar magnetospheres.

In order to reproduce the observed flat radio spectrum, most of recent spectral models of PWNe have simply assumed a broken power-law distribution of accelerated particles at injection without theoretical background for such a distribution \citep[e.g.,][]{Venter&deJager07, deJager08, Zhang+08, Bucciantini+11, Tanaka&Takahara10, Tanaka&Takahara11, Tanaka&Takahara13b, Martin+12, Martin+14, Martin+16, Torres+13a, Torres+13b, Torres+14, Gelfand+15, Tanaka16}.
Other phenomenological ways to account for the radio-emitting particles are introducing two distinct power-law components \citep[e.g.,][]{Atoyan&Aharonian96, deJager+08, Volpi+08, Vorster+13, Olmi+14} or the relativistic Maxwellian with a power-law tail \citep[][]{Fang&Zhang10}, although the latter cannot explain the flat radio spectrum of the Crab Nebula \citep{Zhu+15}.
On the other hand, there are some physically motivated studies for the origin of the flat radio spectrum; diffusive synchrotron radiation model \citep[][]{Fleishman&Bietenholz07}, and source inhomogeneities of the magnetic field and the particle distribution combined with synchrotron cooling effect \citep[][]{Reynolds09} or with adiabatic cooling effect \citep[][]{Tanaka13}.
However, none of them accounts for the broadband spectrum including high energy $\gamma$-rays.
Here, we consider another physical model: stochastic or second-order Fermi acceleration \citep[c.f.,][]{Fermi49, Schlickeiser89}.
Phenomenological stochastic acceleration models have succeeded in reproducing photon spectra of blazars \citep{Asano+14,Kakuwa+15,Asano&Hayashida15},
and Fermi bubbles \citep{Mertsch&Sarkar11,Sasaki+15}. Those successes encourage us to consider the stochastic acceleration in PWNe.

The importance of the stochastic acceleration via turbulences inside PWNe has been already mentioned in some previous studies \citep[][]{Bucciantini+11, Komissarov13, Olmi+14, Porth+16}.
Multi-dimensional numerical studies have showed that PWNe are at a highly turbulent state.
For example, the interaction of the pulsar wind with the SN ejecta induces the Rayleigh-Taylor instability \citep[e.g.,][]{Blondin+01, Porth+14b}.
The anisotropic energy flux of the pulsar wind forms the vortex flow \citep[e.g.,][]{Komissarov&Lyubarsky04, DelZanna+04}, and also the kink instability is expected inside PWNe \citep[e.g.,][]{Begelman98, Porth+14a}.
Polarization observations \citep[e.g.,][]{Bietenholz&Kronberg91, Aumont+10, Moran+13} infer that the Crab Nebula has a significantly turbulent magnetic field structure \citep[c.f.,][]{Shibata+03, Bucciantini+05}.
In addition, in order to explain the observations of the X-ray spectral index and brightness distributions \citep[e.g.,][]{Slane+00, Slane+04, Bamba+10}, the spatial particle transport by the turbulent diffusion has been studied \citep[][]{Tang&Chevalier12, Vorster&Moraal13, Vorster&Moraal14, Porth+16}.
It is natural to consider the effect of the momentum diffusion inside PWNe, since the spatial and momentum diffusions are related to each other \citep[c.f.,][]{Schlickeiser89}.

In this paper, we propose a model of the broadband spectrum of the Crab Nebula, where the radio emission originates in a different component from the pulsar wind itself.
The particles required for the radio component are promptly injected as relics associated with the SN explosion \citep[e.g.,][]{Murase+15}
or continuously supplied from the surrounding SN ejecta as the nebula expands.
In the latter scenario, the SN ejecta filaments penetrating deeply into PWNe \citep[e.g.,][]{Lyutikov03, Komissarov13}
may be mixed with the pulsar wind, and provide the extra particle injection.
The extra particles attributed to the radio emission are stochastically accelerated by turbulence excited via the interaction between the PWN and SN ejecta.
On the other hand, the pulsar wind particles are accelerated at the termination shock and contribute to the emission from optical to X-ray \citep{Kennel&Coroniti84b},
for which the required number flux of the pulsar wind is consistent with the theoretical studies of pulsar magnetospheres.
In our model, the number of the extra particles can largely dominate the number of particles in the nebula,
and the hard spectral distribution is also achieved by the stochastic acceleration.
In order to demonstrate this possibility in this paper,
we show the resultant spectra calculated with a time-dependent model of \citet[][hereafter, TT10]{Tanaka&Takahara10}
with the stochastic acceleration of particles.
Although there are two emission components in our model, for simplicity,
the one-zone treatment in TT10 is adopted.
In Section \ref{sec:Model}, we describe our stochastic acceleration model of PWNe.
Especially, the functional forms of the particles injection to the stochastic acceleration process and of the momentum diffusion coefficient are important ingredients of the present model.
In Section \ref{sec:Application}, we apply the model to the Crab Nebula.
The results are discussed in Section \ref{sec:Discussion} and we conclude the present paper in Section \ref{sec:Conclusions}.

\section{Model}\label{sec:Model}

While the origin of the radio emitting particles is different from the pulsar wind in the present assumption,
a complicated model such as two-zone model, for example, the pulsar wind zone well inside the PWN and the turbulent-acceleration zone near the edge of the PWN, demands increase of model parameters.
Considering the highly disturbed plasma in the MHD simulations such as \citet{Porth+14b, Porth+14a}, a model with distinctively separated two zones
is not necessarily close to the real situation compared to the one-zone treatment.
Here, as the first step to demonstrate the capability of the stochastic acceleration model,
we carry out numerical calculations using our one-zone time-dependent model in TT10.

We consider a PWN as a uniform sphere expanding at a constant velocity $v_{\rm PWN}$ and then the radius of the PWN is $R_{\rm PWN}(t) = v_{\rm PWN} t$.
The emission processes of the PWN are synchrotron radiation and inverse Compton scattering off synchrotron radiation (SSC) and off the cosmic microwave background (CMB).
We ignore the interstellar radiation field of the galactic origin because the Crab Nebula lies far from the Galactic center and disk.
Introducing the magnetic fraction parameter $\eta_{\rm B}$, we assume that the magnetic field inside the PWN $B(t)$ evolves according to
\begin{eqnarray}\label{eq:BfieldEvolution}
	\frac{4 \pi}{3} R^3_{\rm PWN}(t) \frac{B^2(t)}{8 \pi}
	& = &
	\eta_{\rm B} \int^t_0 L_{\rm spin}(t)
	\equiv
	\eta_{\rm B} E_{\rm rot}(t),
\end{eqnarray}
where $L_{\rm spin}(t) = L_0 (1 + t/\tau_0)^{-(n+1)/(n-1)}$ is the spin-down power of the central pulsar, $n$ is braking index, and $\tau_0$ is spin-down time.
The initial rotational energy of the pulsar is $L_0 \tau_0$.
Although different evolution models of $R_{\rm PWN}(t)$ and $B(t)$ are used in some papers \citep[c.f.,][]{Torres+14}, in order to compare some parameters directly to our past studies, we adopt the above basic assumptions that are the same as our past models.

The evolution of the particle energy distribution in the PWN $N(\gamma, t)$ including the stochastic acceleration is described by the Fokker-Planck equation
\begin{eqnarray}\label{eq:FPEquation}
	\frac{\partial}{\partial t}      N(\gamma, t)
	& + &
	\frac{\partial}{\partial \gamma}
	\left[
		\left(
			\dot{\gamma}_{\rm cool}(\gamma, t)
			-
			\gamma^2 D_{\gamma \gamma}(\gamma, t) \frac{\partial}{\partial \gamma} \frac{1}{\gamma^2}
		\right) N(\gamma, t)
	\right]
	\nonumber \\
	& = &
	Q_{\rm PSR}(\gamma, t)
	+
	Q_{\rm ext}(\gamma, t),
\end{eqnarray}
where $\gamma$ is the Lorentz factor of electrons/positrons.
For the cooling term $\dot{\gamma}_{\rm cool}(\gamma, t)$, we use the same expression as our past studies, which include adiabatic cooling $\dot{\gamma}_{\rm ad}(\gamma,t)$, synchrotron radiation $\dot{\gamma}_{\rm syn}(\gamma,t)$, and inverse Compton scattering $\dot{\gamma}_{\rm IC}(\gamma)$.
We introduce the two injection terms $Q_{\rm PSR}(\gamma, t)$ and $Q_{\rm ext}(\gamma, t)$.
The term $Q_{\rm PSR}(\gamma, t)$ represents the injection from the central pulsar.
The synchrotron radiation from optical through X-rays \citep[c.f.,][]{Kennel&Coroniti84b} is attributed to the particles originated from $Q_{\rm PSR}$.
We call those particles injected from the pulsar the `X-ray-emitting particles' below.
The extra particle injection $Q_{\rm ext}(\gamma, t)$ is the source of the radio-emitting particles.
The particles are stochastically accelerated by the term expressed with the diffusion coefficient in energy space $D_{\gamma \gamma}$.

The particle injection from the pulsar $Q_{\rm PSR}(\gamma, t)$ is assumed to have a `single' power-law distribution characterized by the three parameters, the minimum  $\gamma_{\rm min}$ and maximum $\gamma_{\rm max}$ Lorentz factors and the power-law index $p$,
\begin{eqnarray}\label{eq:PSRinjection}
	Q_{\rm PSR}(\gamma, t)
	& = &
	\dot{n}_{\rm PSR}(t) \gamma^{-p}
	H(\gamma - \gamma_{\rm min}) H(\gamma_{\rm max} - \gamma),
\end{eqnarray}
where $H$ is Heaviside's step function.
The normalization $\dot{n}_{\rm PSR}(t)$ is set to satisfy $\int d \gamma Q_{\rm PSR}(\gamma, t) \gamma m_{\rm e} c^2 d \gamma = \eta_{\rm e} L_{\rm spin}(t)$ with $\eta_{\rm e} \lesssim 1$.
We adopt the value of $\gamma_{\rm min} \sim 10^6$, which is closely related with the bulk Lorentz factor of the pulsar wind \citep[c.f.,][]{Kennel&Coroniti84b}, and then the particle number flux of the pulsar wind becomes consistent with the theoretical expectation \citep{Daugherty&Harding82,Hibschman&Arons01,Medin&Lai10,Timokhin&Harding15}.
Another difference from our past studies is that we set $\eta_{\rm e} < 1 - \eta_{\rm B}$.
The rest of $L_{\rm spin}$, i.e., $(1 - \eta_{\rm B} - \eta_{\rm e}) L_{\rm spin}$, is considered to be converted into an energy of turbulence that accelerates the radio-emitting particles.
The turbulent energy should be non-negligible amount because the radio-emitting particles have a significant fraction of the injected rotational energy $E_{\rm rot}(t)$ \citep[c.f.,][]{Kennel&Coroniti84b, Komissarov13}.

The other injection term $Q_{\rm ext}(\gamma, t)$ accounts for the particles that have another origin, for example, mixing of line-emitting plasmas from the SN ejecta filaments \citep[e.g.,][]{Lyutikov03, Komissarov13}.
For the radio-emitting particles,
only a fraction of the plasmas, e.g., the supra-thermal tail of the momentum distribution, is injected to the stochastic acceleration process, and the injection energy $\gamma_{\rm inj}$ would be mildly relativistic regime.
We have tested three different values of $\gamma_{\rm inj} =$ 1, 2, and 10, and we have not found a significant difference
in the resultant particle spectra. The slight difference of the normalization due to a different $\gamma_{\rm inj}$ can be adjusted by the normalization parameters $N_{\rm imp}$ or $\dot{n}_{\rm cont}$.
We consider two forms of time-dependence of $Q_{\rm ext}(\gamma, t)$, the impulsive and the continuous injections, and we express them
\begin{eqnarray}\label{eq:SNinjection}
	Q_{\rm ext}(\gamma, t)
	& = &
	\left\{
   	\begin{array}{ll}
	Q_{\rm imp}(\gamma, t)
	\equiv
	N_{\rm imp} \delta(t) \delta(\gamma - \gamma_{\rm inj}), \\
	Q_{\rm cont}(\gamma, t)
	\equiv
	\dot{n}_{\rm cont} \left( t / t_{\rm age} \right)^s \delta(\gamma - \gamma_{\rm inj}), \\
   	\end{array} \right.
\end{eqnarray}
respectively.

For the case of the impulsive injection $Q_{\rm imp}(\gamma, t)$, we consider the possibility that the plenty of the particles are supplied at the very initial phase of the PWN evolution.
At the very early stage ($\sim$ weeks to months), the PWN would be thick to pair creation process \citep[c.f.,][]{Murase+15}.
Although those secondary particles are immediately cooled by the expansion of the PWN, some fraction of them will be injected to the acceleration process by turbulence.
The number $N_{\rm imp}$ is the normalization parameter in the impulsive injection case.
The impulsive approximation (prompt injection at $t=0$) implies that
the duration of the particle injection should be significantly shorter than
the time-scales $(\tau_{\rm acc,m}, \tau_{\rm turb})$ that are parameters of the energy diffusion coefficient defined in Equation (\ref{eq:MomDiffCoefficient}).

On the other hand, for the case of the continuous injection $Q_{\rm cont}(\gamma, t)$, we consider the possibility of continuous particle injection from the SN ejecta.
The neutral particles in the line-emitting filament can penetrate into the PWN, and be injected into the turbulent acceleration process
by photoionization \citep[c.f.,][]{Morlino+15}.
Alternatively, the Rayleigh-Taylor turbulence disturbs the contact discontinuity between the PWN and the shocked ejecta.
In such a region, the stochastic acceleration may work.
In this model, the model parameters are the normalization $\dot{n}_{\rm cont}$ defined with the present age of the PWN $t_{\rm age}$,
the temporal index $s$, and $\gamma_{\rm inj}$.
If the injected particle number is proportional to the SN ejecta mass swept up by the PWN,
$\dot{M}_{\rm sw} = 4 \pi R^2_{\rm PWN}(t) v_{\rm PWN} \rho_{\rm ej}$, where $\rho_{\rm ej}$ is the density of SN ejecta,
we can adopt $s = 2$ as a fiducial value for constant $\rho_{\rm ej}$.
The value $s = 2$ may be an almost maximum value, since, in the real situation, $\rho_{\rm ej}$ would decrease with the time
from the SN explosion and/or with the distance from the explosion center \citep[c.f.,][]{Blondin+01, vanderSwaluw+01}.

The term with $D_{\gamma \gamma}$ in Equation (\ref{eq:FPEquation}) represents the stochastic acceleration and diffusion in the energy space.
The magnetic energy is much smaller than the particle energy in PWNe, so that
the Alfv\'en wave seems not to be responsible for the energy source of the radio-emitting particles.
The alternative candidates for the particle scatterer may be the hydrodynamically induced eddy
or the fast wave, which can retain its energy density larger than the background magnetic energy density.
When the energy of the fast wave cascades to significantly small scales (i.e. large wave number parallel to the magnetic field $k_\parallel$),
relativistic particles can resonate with such short waves following $\omega-k_\parallel c \mu=n_{\rm F} \Omega$,
where $\omega$ is the wave frequency, $\mu$ is the cosine of the pitch angle,
and $\Omega$ is the gyrofrequency of the particle.
The case of $n_{\rm F}=0$, so-called transit-time damping (TTD),
corresponds to the resonant interaction with magnetic mirror force,
and the gyroresonance interaction is represented by the mode of $n_{\rm F}=\pm 1$.
In such resonant interactions with the fast wave following
the power-law wave energy distribution $W(k) \propto k^{-q}$,
the energy diffusion coefficient is proportional to $\gamma^{q}$
\citep[e.g.][]{Schlickeiser&Miller98}.
On the other hand, when the cascade is damped at a large scale or a large-scale eddy is responsible for the acceleration,
the non-resonant interaction with the compressible fluid leads to
the hard-sphere formula $D_{\gamma \gamma} \propto \gamma^2$ \citep[e.g.][]{Ptuskin88},
where the acceleration time-scale $\sim \gamma^2/D_{\gamma \gamma}$ is energy-independent.

We phenomenologically describe the functional form of the diffusion coefficient as
\begin{eqnarray}\label{eq:MomDiffCoefficient}
	D_{\gamma \gamma}(\gamma, t)
	& \equiv &
	\frac{\gamma^2_{\rm min}}{2 \tau_{\rm acc,m}}
	\left( \frac{\gamma}{\gamma_{\rm min}} \right)^q
	\exp \left( - \frac{t     }{  \tau_{\rm turb}} \right)
	\exp \left( - \frac{\gamma}{\gamma_{\rm cut}} \right),
\end{eqnarray}
where the parameters are the acceleration time-scale $\tau_{\rm acc,m}$ at $\gamma=\gamma_{\rm min}$.
(note that $\gamma_{\rm min} \neq \gamma_{\rm inj}$), the duration time-scale of the turbulence acceleration $\tau_{\rm turb}$,
and the cut-off Lorentz factor $\gamma_{\rm cut}$, which is artificially introduced to avoid the acceleration of X-ray-emitting particles
in our one-zone treatment.
The acceleration time becomes longer at higher energy for $q<2$ as $t_{\rm acc}=\gamma^2 /(2 D_{\gamma \gamma}) \propto \gamma^{2-q}$.
The duration $\tau_{\rm turb}$ is interpreted as a time-scale at which the feedback process starts to decay the turbulence,
or the interaction between the SN ejecta and pulsar wind becomes insufficient to induce the turbulence as the PWN and ejecta expand.
Here, for simplicity, we assume that $D_{\gamma \gamma}$ is almost constant for $t<\tau_{\rm turb}$.
The relevance of this simple assumption will be discussed in section \ref{sec:Discussion}.
Note that the spatial diffusion coefficient is inversely proportional to the momentum diffusion coefficient in general.

\section{Application}\label{sec:Application}

%
\begin{table}[!t]
\caption{
	Summary of the parameters for the calculations in Figures \ref{fig:SimpleDgg}, \ref{fig:model1}, \ref{fig:model2} corresponding to the simple continuous acceleration model, Models 1 and 2, respectively.
}
\label{tbl:Parameters}
\begin{tabular}{ccccc}
	Symbol  &
	TT10    &
	Simple  &
	Model 1 &
	Model 2 \\
\hline
\multicolumn{5}{c}{Fixed Parameters} \\
\hline
$P$(ms)                    & \multicolumn{4}{c}{33.1}                   \\
$\dot{P}$(s s$^{-1}$)      & \multicolumn{4}{c}{4.21 $\times 10^{-13}$} \\
$n$                        & \multicolumn{4}{c}{2.51}                   \\
$t_{\rm age}$(yr)          & \multicolumn{4}{c}{950}                    \\

$v_{\rm PWN}$(km s$^{-1}$) & \multicolumn{4}{c}{1800}                   \\
$d$(kpc)                   & \multicolumn{4}{c}{2.0}                    \\
%
%
%
$\eta_{\rm B}$             & \multicolumn{4}{c}{5.0 $\times 10^{-3}$}   \\
$\gamma_{\rm inj}$                        & $-$ & 2        & 2      & 2      \\
$s$                                       & $-$ & 2        & 2      & $-$    \\
\hline
\multicolumn{5}{c}{Fitted Parameters}    \\
\hline
$\eta_{\rm e}$                            & 1   & 0.35     & 0.6    & 0.6    \\
$p$                                       & 2.5 & 2.4      & 2.5    & 2.5    \\
$\gamma_{\rm min}$($10^{5}$)
\footnotemark[1]                          & 6.0 & 4.0      & 4.0    & 4.0    \\
$\gamma_{\rm max}$($10^{9}$)              & 7.0 & 7.0      & 7.0    & 7.0    \\
$N_{\rm imp}$($10^{50}$)                  & $-$ & $-$      & $-$    & 20     \\
$\dot{n}_{\rm cont}$($10^{40}~{\rm s}^{-1}$)    & $-$ & 10       & 600    & $-$    \\
$\tau_{\rm acc,m}$(yr)                    & $-$ & 950      & 25     & 50     \\
$\tau_{\rm turb}$(yr)                    & $-$ & $\infty$ & 250    & 300    \\
$\gamma_{\rm cut}$                      & $-$ & $\infty$ & $10^5$ & $10^5$ \\
$q$                                       & $-$ & 5/3      & 2      & 2      \\
\hline
\multicolumn{5}{c}{Dependent Parameters} \\
\hline
$N_{\rm PSR}$($10^{50}$)                  & 30  & 0.12     & 0.23   & 0.23   \\
$N_{\rm ext}$($10^{50}$)                  & $-$ & 10       & 600    & 20     \\

\end{tabular}

\footnotetext[1]{
	Corresponding to $\gamma_{\rm b}$ for the value of TT10.
}
\end{table}

Here, the model is applied to the Crab Nebula.
The same observational properties of the Crab Nebula as TT10 is adopted; the period $P = 33.1$ ms, its derivative $\dot{P} = 4.21 \times 10^{-13}~{\rm s~s^{-1}}$, the braking index $n = 2.51$, the age $t_{\rm age} = 950$ yr, and the expansion velocity $v_{\rm PWN} = 1800~{\rm km~s^{-1}}$.
Assuming the moment of inertia of the pulsar as $10^{45}~\mbox{g}~\mbox{cm}^2$, those values imply $L_0 \simeq 3.4 \times 10^{39}~\mbox{erg}~\mbox{s}^{-1}$
and $\tau_0 \simeq 700$ yr.
In addition, we fix the magnetic fraction $\eta_{\rm B} = 5 \times 10^{-3}$ and then the present magnetic field strength of 85 $\mu$G is also the same as TT10.

We adjust the remaining parameters in $Q_{\rm PSR}(\gamma, t)$ (Equation (\ref{eq:PSRinjection})), $Q_{\rm ext}(\gamma, t)$ (Equation (\ref{eq:SNinjection})), and $D_{\gamma \gamma}$ (Equation (\ref{eq:MomDiffCoefficient})) to reproduce the entire spectrum.
The four parameters for $Q_{\rm PSR}(\gamma, t)$ ($\gamma_{\rm min},~\gamma_{\rm max},~p$ and $\eta_{\rm e}$) hardly affect the radio emission and we choose these parameters after the other parameters are fitted.
For the extra injection $Q_{\rm ext}(\gamma, t)$, we consider both the continuous (Model 1) and impulsive (Model 2) injection cases.
The calculated radio flux is exactly proportional to the normalizations $N_{\rm imp}$ or $\dot{n}_{\rm cont}$, while $\gamma_{\rm inj}$ makes only a little change of the normalization of the radio flux as long as $\gamma_{\rm inj} \le 10 \ll \gamma_{\rm min}$.
We set $\gamma_{\rm inj} = 2$ and use $N_{\rm imp}$ or $\dot{n}_{\rm cont}$ to fit the normalization of the radio flux.
We adopt $s = 2$ for the continuous injection case as mentioned in the previous section.
The parameters for $D_{\gamma \gamma}$ ($\tau_{\rm acc,m},~\tau_{\rm turb},~\gamma_{\rm cut}$, and $q$) are main points in this paper to reproduce the flat radio spectrum of the Crab Nebula.

As will be found below, the broadband spectrum of the Crab Nebula alone is reproduced with various combinations of the non-trivial four parameters of $D_{\gamma \gamma}$.
We require the other conditions to constrain these parameters.
First of all, we impose that the energy content of the Crab Nebula dose not exceed the rotational energy supplied from the Crab pulsar.
In contrast to TT10, this condition for the energetics is not always satisfied because of the stochastic acceleration,
whose parameters are not regulated by the energy budget in our numerical code.
In section \ref{sec:Simple}, we show a simple model ($\tau_{\rm turb}=\infty$, $\gamma_{\rm cut}=\infty$)
that reproduces the broadband spectrum, but does not explain the observed flux evolution.
The results provide us a hint how to constrain the parameters for the stochastic acceleration.
In section \ref{sec:HardSphereModels}, we demonstrate the possible model improvement by introducing additional parameters such as $\tau_{\rm turb}$ and $\gamma_{\rm cut}$.
All the parameters are summarized in Table \ref{tbl:Parameters}.
Because parameter sets which reproduce the observations are not unique, we summarize the general conclusions of the present application to the Crab Nebula in section \ref{sec:Summary}.

\subsection{Simple Model with Continuous Injection and Acceleration}\label{sec:Simple}

%
\begin{figure*}
\begin{center}
\includegraphics[scale=0.6]{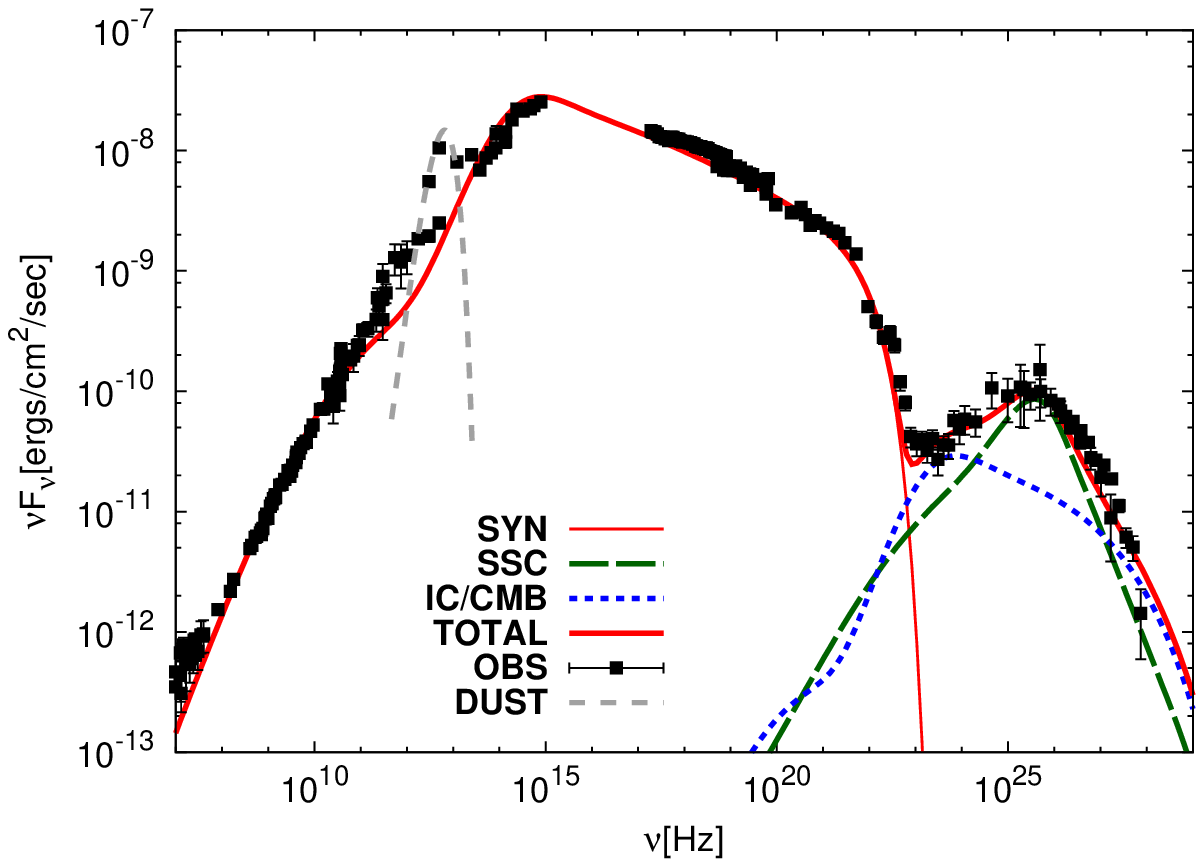}
\includegraphics[scale=0.6]{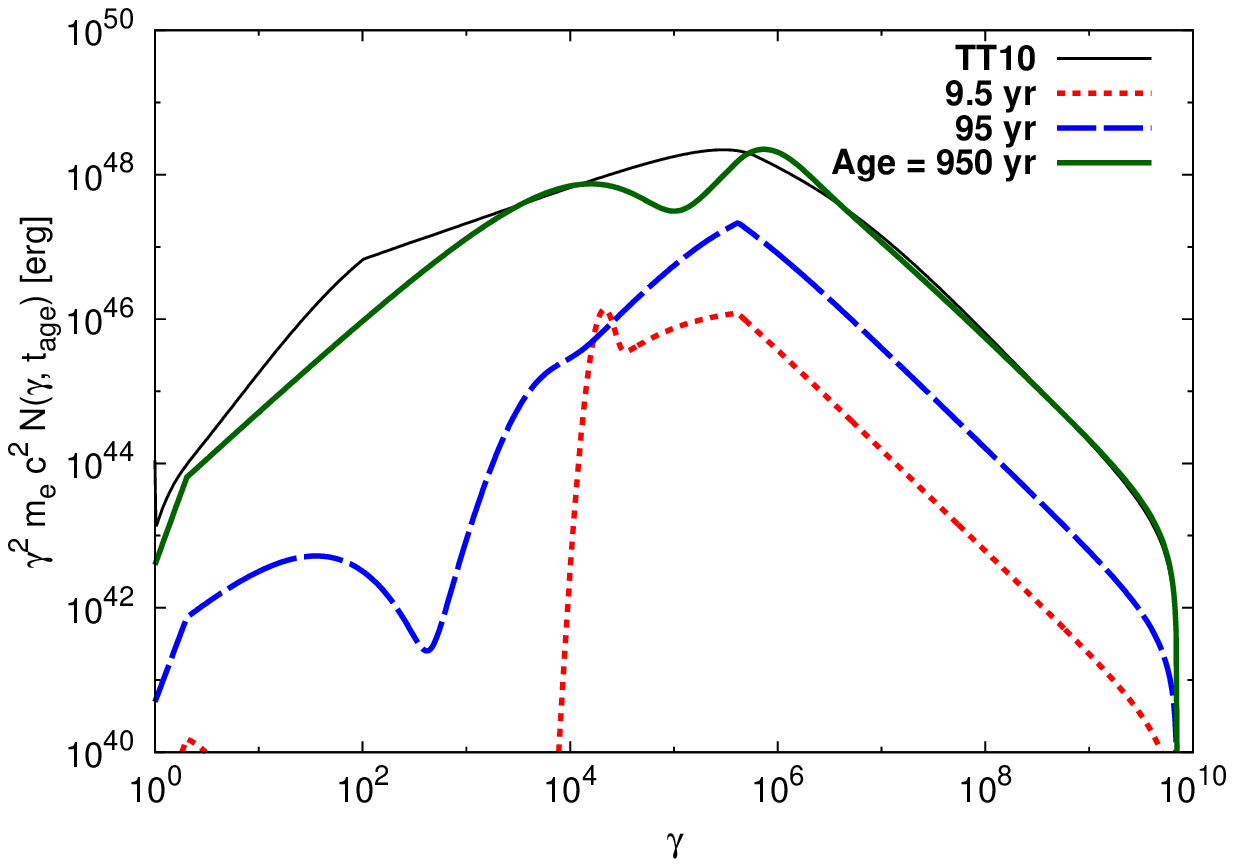} \\
\includegraphics[scale=0.6]{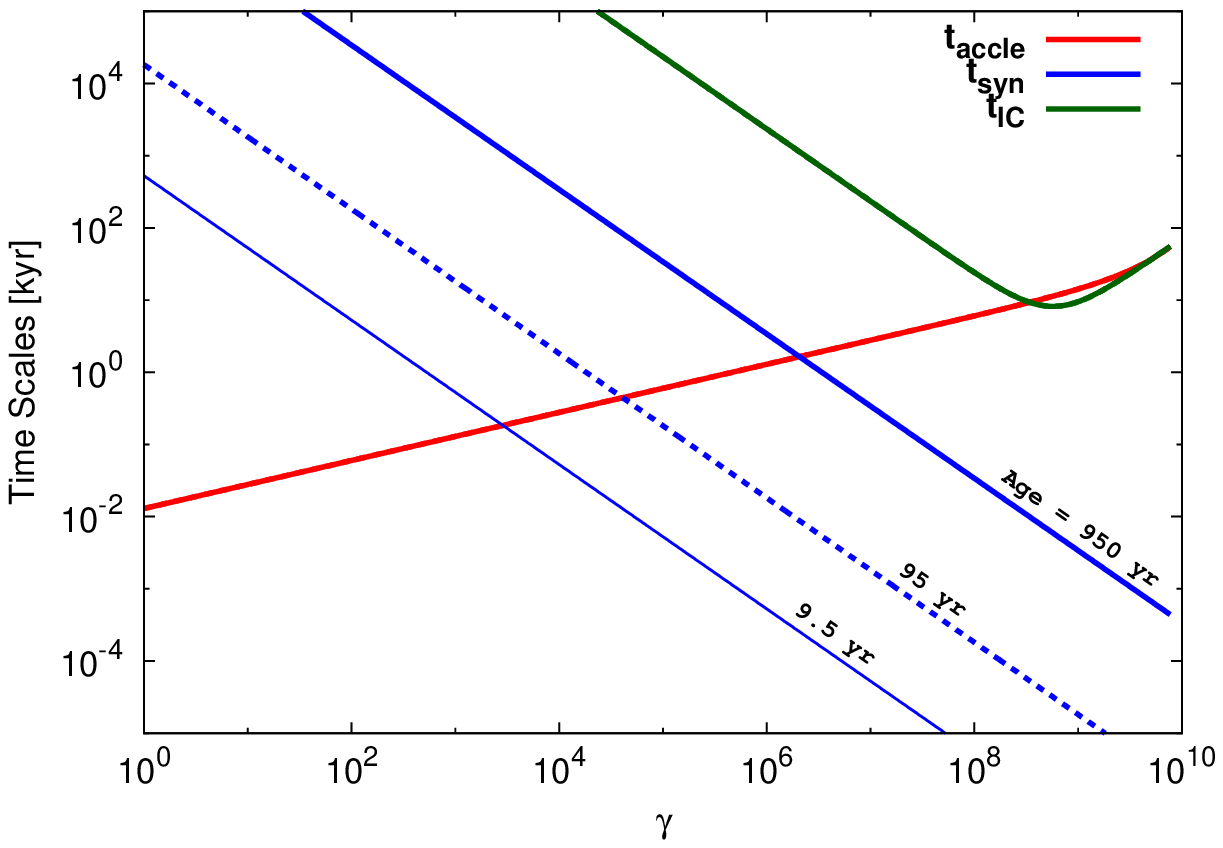}
\includegraphics[scale=0.6]{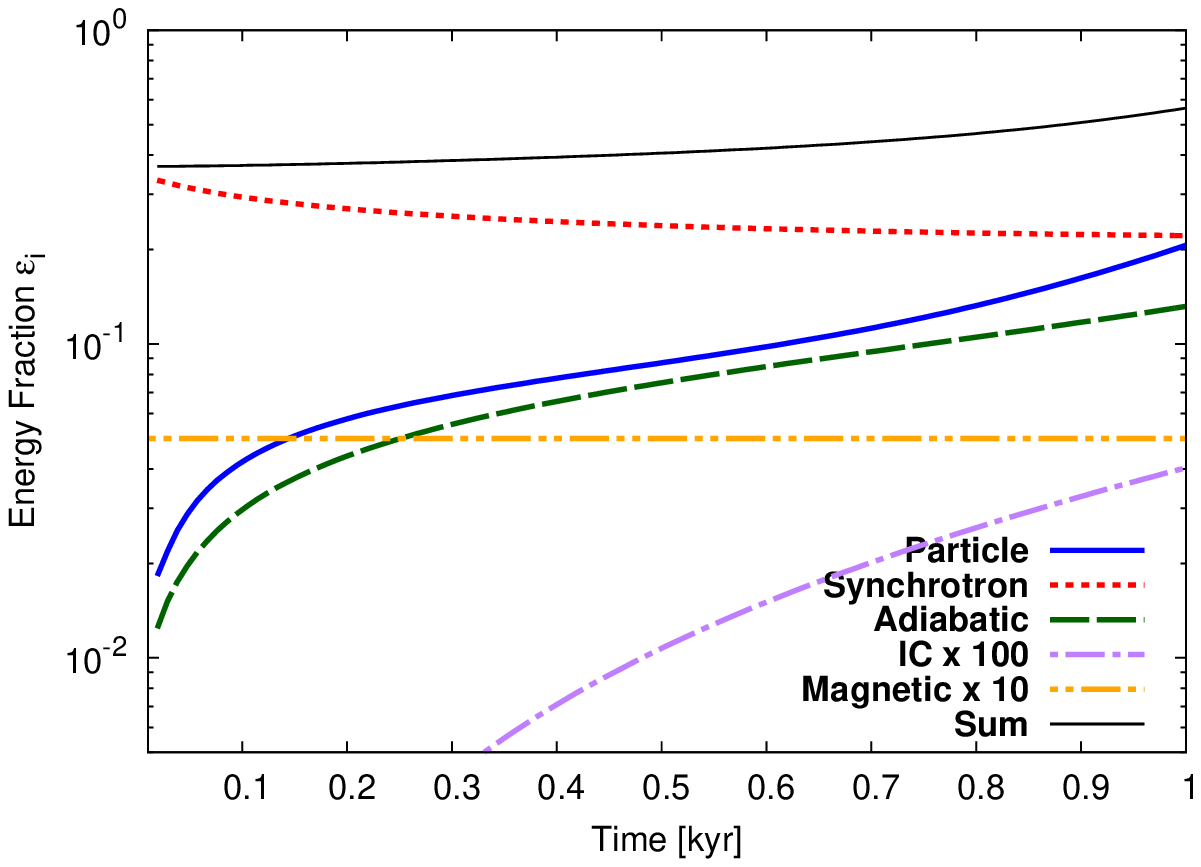}
\end{center}
\caption{
\label{fig:SimpleDgg}
	The results of the simple Kolmogorov-like case
	(see parameters in Table \ref{tbl:Parameters}).
	Top-left: the current photon spectrum with the observational data (\citet{Baldwin71, Baars+77, Macias-Perez+10} for radio, \citet{Ney&Stein68, Grasdalen79, Green+04, Temim+06} for IR, \citet{Kuiper+01} for X-rays, and \citet{Aharonian+06, Albert+08, Abdo+10} for $\gamma$-rays).
	Top-right: evolution of the particle spectra from $10^{-2} t_{\rm age} =$9.5 yr (dotted red) to $t_{\rm age} =$950 yr (thick green) with the result of TT10 at $t_{\rm age}$ (thin black).
	Bottom-left: the time-scales of the acceleration $t_{\rm acc}(\gamma, t) = \gamma^2 / 2 D_{\gamma \gamma}(\gamma, t)$ (positive slope red line), synchrotron cooling $t_{\rm syn} = \gamma / \dot{\gamma}_{\rm syn}$ (negative slope blue lines), and the inverse Compton cooling $t_{\rm IC} = \gamma / \dot{\gamma}_{\rm IC}$ (concave upward green line) for different ages.
	Only $t_{\rm syn}$ changes with time in this case (thin, dotted and thick lines correspond to $10^{-2} t_{\rm age}$, $10^{-1} t_{\rm age}$ and $t_{\rm age}$, respectively).
	Bottom-right: evolution of the fractional energy of the particles $\int d \gamma \gamma m_{\rm e} c^2 N(\gamma, t)$ (thick blue), magnetic field $\eta_{\rm B} E_{\rm rot}(t)$ (dot-dot dashed yellow, ten times amplified), of the radiated and wasted ones by synchrotron (dotted red), inverse Compton (dot-dashed purple, $10^2$ times amplified) and adiabatic (dashed green) cooling $\int^t_0 d t'\int d \gamma |\dot{\gamma}_i(\gamma, t')| m_{\rm e} c^2 N(\gamma, t')$ ($i =$ syn, IC, ad) and of the sum of all of them (thin black).
	All the components are divided by $E_{\rm rot}(t)$, which is increasing with time.
}
\end{figure*}
\begin{figure}
\includegraphics[scale=0.7]{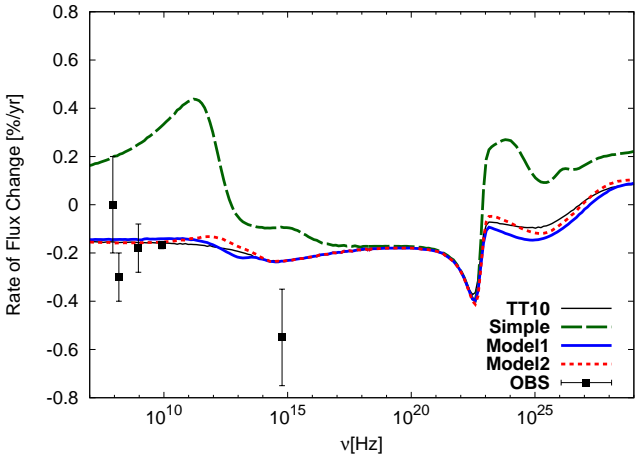}
\caption{
	The present rate of flux change from radio to $\gamma$-rays for the different models with the observations (\citet{Vinyaikin07} for radio and \citet{Smith03} for optical).
	Thin black line is TT10.
	The simple continuous acceleration model (see section \ref{sec:Simple}) corresponds to Green dashed line.
	Thick blue and dotted red lines are the results of Models 1 and 2 in section \ref{sec:HardSphereModels}, respectively.
\label{fig:FluxEvolution}
}
\end{figure}

As the first step, we adopt a simple model with the Kolmogorov-like index $q=5/3$, continuous injection ($Q_{\rm ext} = Q_{\rm cont}$),
and continuous acceleration ($\tau_{\rm turb}=\infty$, $\gamma_{\rm cut}=\infty$).
In addition, here we set the acceleration time-scale as $\tau_{\rm acc,m}=t_{\rm age}$ in order to accelerate the radio-emitting particles from $\gamma_{\rm inj}$ to $\gamma_{\rm min}$ within the age of system.
In this case, the parameter to be adjusted is only $\dot{n}_{\rm cont}$.
As shown in Figure \ref{fig:SimpleDgg},
the parameter value $\dot{n}_{\rm cont} = 1.0 \times 10^{41}~{\rm s^{-1}}$ marginally reproduce the current broadband spectrum of the Crab Nebula.
In this Kolmogorov-like case, the turbulence acceleration hardly affects the evolution of the particles injected above $\gamma_{\rm min}$ (compare the acceleration and cooling time-scales for high energy particles in the bottom-left panel of Figure \ref{fig:SimpleDgg}),
so that the spectrum of the X-ray-emitting particles almost coincides with the result of TT10.
Note that the required small $\eta_{\rm e} \sim 0.35$ compared to TT10 is consistent with the energy budget
for the turbulence responsible for the stochastic acceleration (the bottom-right panel of Figure \ref{fig:SimpleDgg}).
As seen in the top-left panel of Figure \ref{fig:SimpleDgg}, the continuously injected particles of $\gamma_{\rm inj} = 2$ are stochastically accelerated and forms the radio-emitting particles which have a peak at $\gamma \sim 10^4$ at $t=t_{\rm age}$ (thick green).
Although the particle spectrum with the two peaks at $t_{\rm age}$ seems significantly different from the result of TT10 (thin black),
the broadband emission spectrum seems not far different from the observations (the top-left panel of Figure \ref{fig:SimpleDgg}).

The observed spectrum at around $10^{12} - 10^{13}$ Hz (we call `the infrared bump' below) contains emission from the dust formed by the SN explosion \citep[c.f.,][]{Temim+12, Owen&Barlow15} and then we overlay the 80 $K$ blackbody spectrum in the broadband spectrum just for reference.
The emission at the frequencies below the infrared bump is synchrotron emission from the stochastically accelerated particles.
We find that the resultant curved spectrum rather than a power-law one does not contradict the flat radio spectrum.
The spectral component above the infrared bump to $\sim 10^{23}$ Hz is due to synchrotron emission from the X-ray-emitting particles accelerated at the termination shock $Q_{\rm PSR}$.
The $\gamma$-ray emission above $10^{23}$ Hz is mostly from the inverse Compton scattering off the radio synchrotron photons by the X-ray-emitting particles.

In this continuous acceleration model without energy cut off ($\tau_{\rm turb} = \infty$, $\gamma_{\rm cut} = \infty$),
the parameter set in Figure \ref{fig:SimpleDgg} is almost the unique one to reproduce the observed broadband spectrum.
The energy-dependent acceleration ($q = 5/3 < 2$) and the acceleration time of $\tau_{\rm acc,m} \sim t_{\rm age}$ are required to avoid accelerating particles of $\gamma \ge \gamma_{\rm min}$ in our one-zone treatment.
The energy independent acceleration ($q = 2$) or shorter acceleration time ($\tau_{\rm acc,m} < t_{\rm age}$) definitely affects the particle spectrum for $\gamma \ge \gamma_{\rm min}$, which leads to a harder X-ray spectrum than observed one.
In addition, for $q < 2$, the increasing injection rate $s = 2$ is essential to form the smooth curved electron spectrum that agrees with the flat radio spectrum.
If we adopt $s=0$, the radio spectrum becomes too hard.

The bottom-right panel of Figure \ref{fig:SimpleDgg} shows evolution of energetics inside the Crab Nebula.
The sum of the particle, magnetic, radiated and adiabatically cooled energies (thin black) is about a half of the injected rotational energy $E_{\rm rot}(t)$ and then this is consistent with $\eta_{\rm B} + \eta_{\rm e} \approx 0.35$.
The rest of $\sim 0.65 E_{\rm rot}(t)$ is interpreted to be stored as the kinetic energy of the turbulent flow and a fraction of it has been used to accelerate the radio-emitting particles already.
However, under the assumption of $\tau_{\rm turb}=\infty$, the particle energy (thick blue) keeps increasing and will exceed $E_{\rm rot}(t)$ for $t \gg t_{\rm age}$.
A finite value of $\tau_{\rm turb}$ as a feedback process from the particles to the turbulence is required to follow the evolution beyond $t_{\rm age}$.

We have another strong motivation to introduce $\tau_{\rm turb}$.
Figure \ref{fig:FluxEvolution} shows the calculated present flux evolution rate with the observations in radio and optical.
The current radio observations clearly show that the flux decreases with time, while our continuous acceleration model (dashed green) predicts increase of the radio flux, because the radio-emitting particles are being accelerated even at present.
We attempt models with finite $\tau_{\rm turb}$ and $\gamma_{\rm cut}$ to reproduce the flux decrease of the radio emission below.
Note that the optical observation would contain potential uncertainty and may not show the secular decrease in contrast to the radio observation \citep[][]{Smith03}.
Actually, the X-ray and $\gamma$-ray variabilities are about an order of magnitude larger, and even increasing and decreasing has been observed \citep[][]{Wilson-Hodge+11}.
Such large variabilities would reflect the contribution from the small emission region close to the termination shock, while the radio flux evolution discussed in Figure \ref{fig:FluxEvolution} is secular evolution of flux from the entire nebula.

\subsection{Hard Sphere Models with Finite Duration of Acceleration}\label{sec:HardSphereModels}

%
\begin{figure*}
\begin{center}
\includegraphics[scale=0.6]{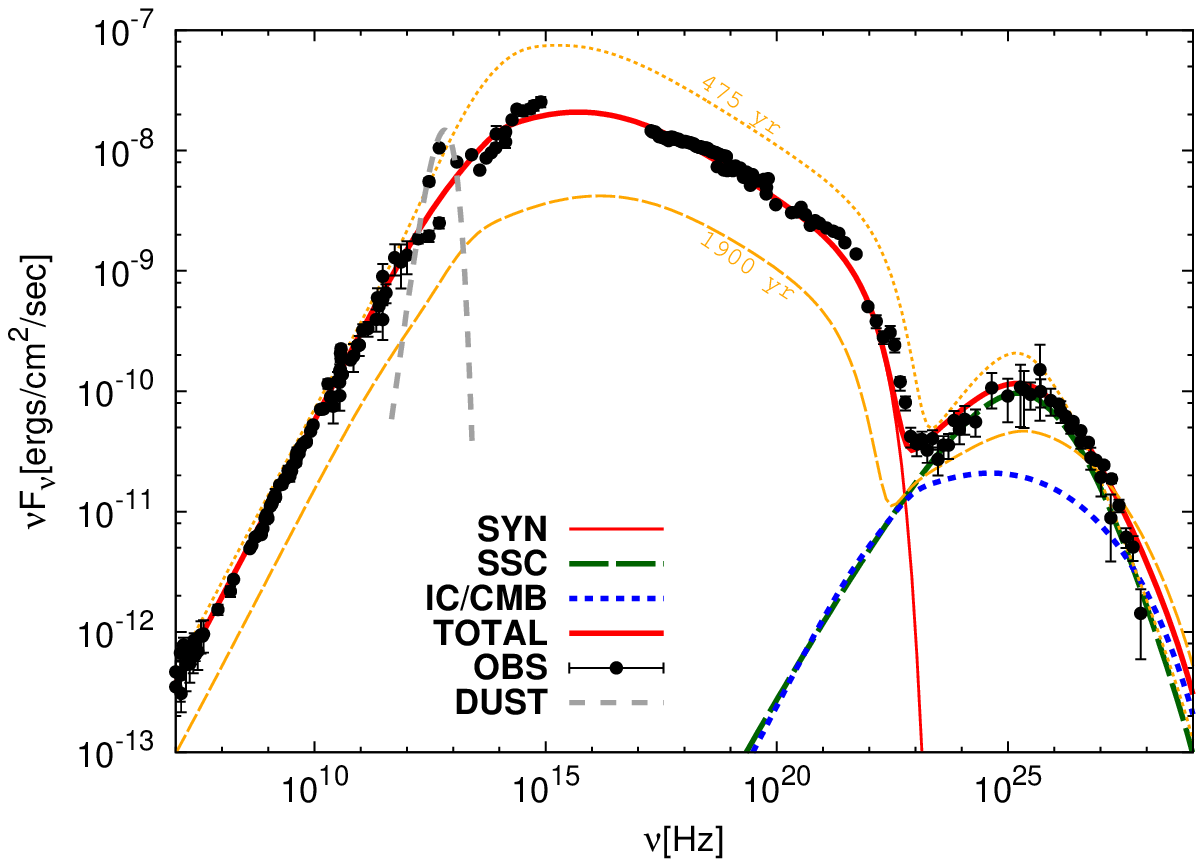}
\includegraphics[scale=0.6]{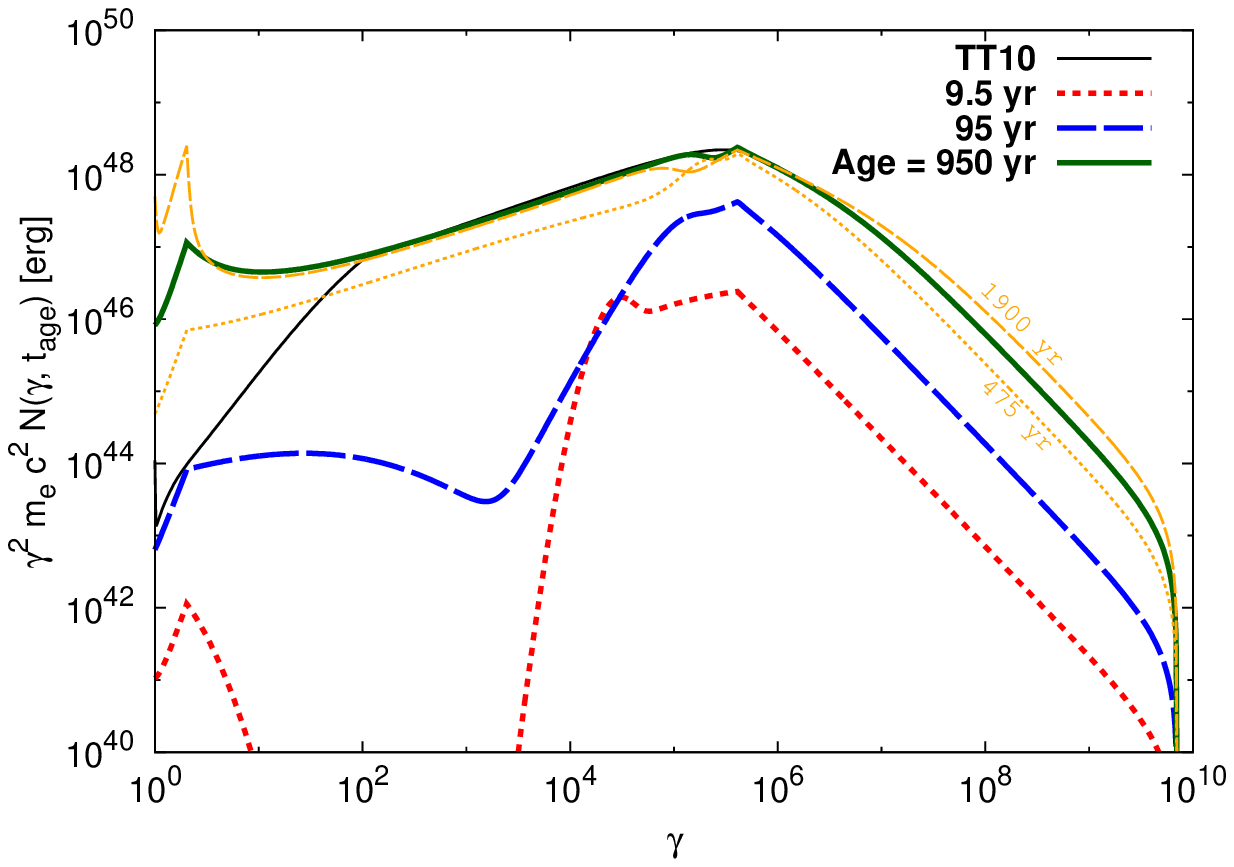} \\
\includegraphics[scale=0.6]{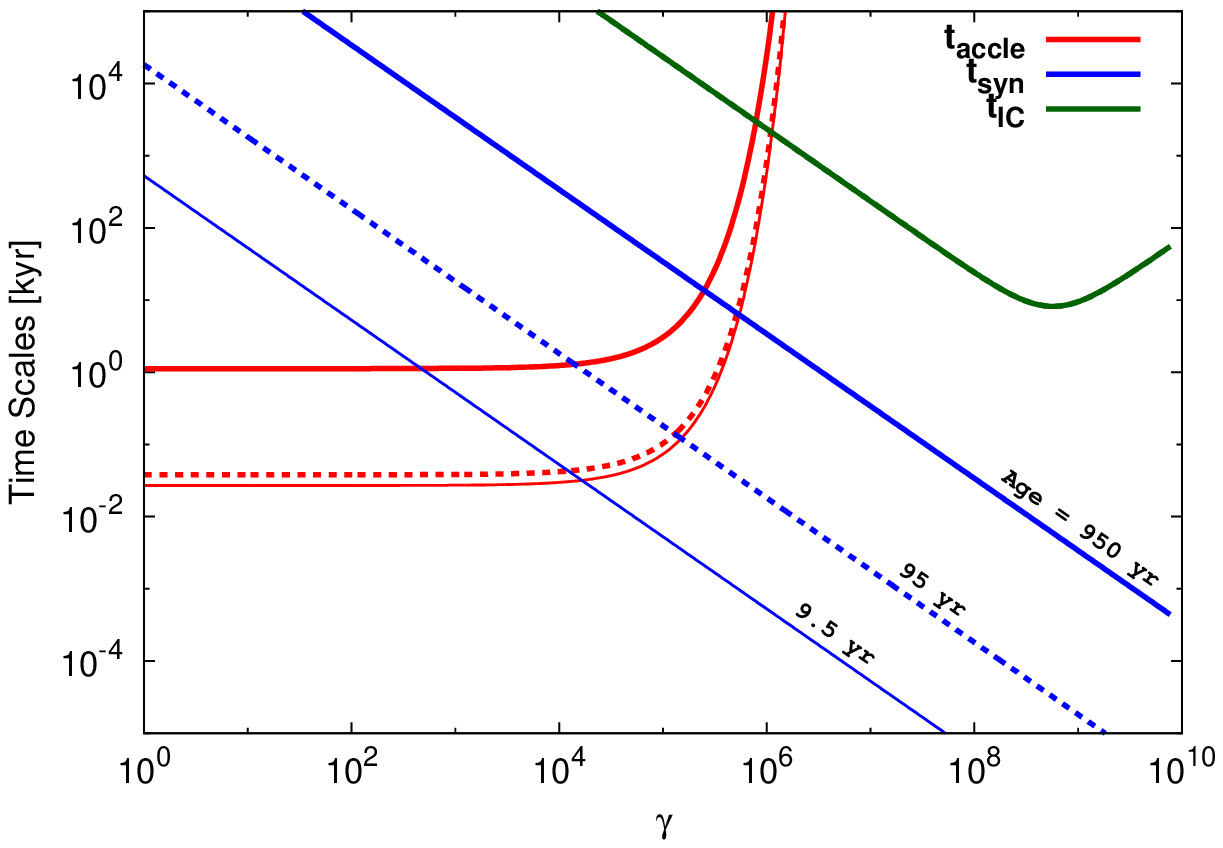}
\includegraphics[scale=0.6]{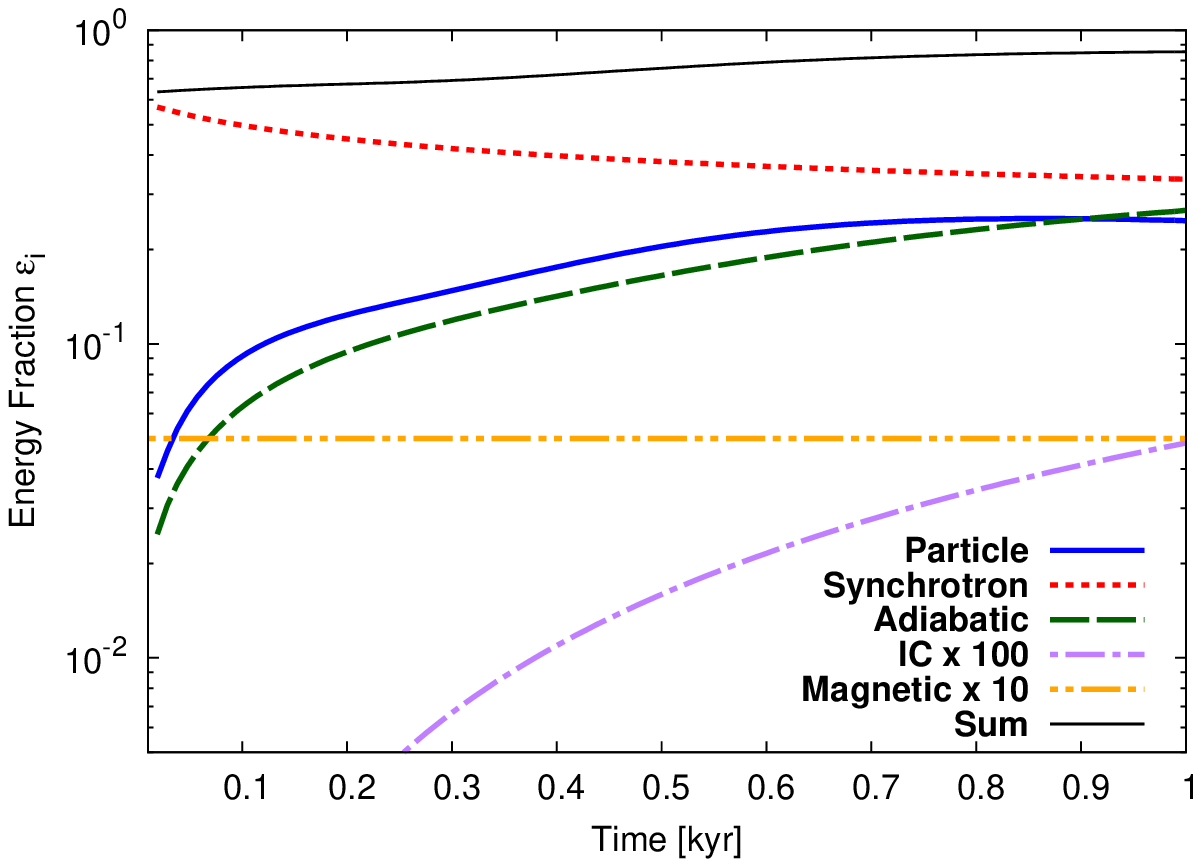}
\end{center}
\caption{
	Calculation results of Model 1: the current photon spectrum (top-left), evolution of the particle spectra (top-right), the time-scales (bottom-left), evolution of the fractional energy (bottom-right).
	See the caption in Figure \ref{fig:SimpleDgg} for details, while, in top-left and -right panels, the photon and particle spectra at $1/2 t_{\rm age} =$ 475 yr (dotted yellow) and $2 t_{\rm age} =$ 1900 yr (dashed yellow) are added (discussed in section \ref{sec:Properties}).
	The dotted and dashed yellow lines in the top-left panels are the photon spectra at $1/2 t_{\rm age} =$ 475 yr and $2 t_{\rm age} =$ 1900 yr, respectively.
	The dashed yellow line in the top-right panel are the particle spectra at $10 t_{\rm age} =$ 9.5 kyr.
	Model 1 adopts $Q_{\rm ext} = Q_{\rm cont}$ and finite values of $\tau_{\rm turb}$ and $\gamma_{\rm cut}$ in Equation (\ref{eq:MomDiffCoefficient}).
	The used parameters are summarized in Table \ref{tbl:Parameters}.
\label{fig:model1}
}
\end{figure*}
\begin{figure*}
\begin{center}
\includegraphics[scale=0.6]{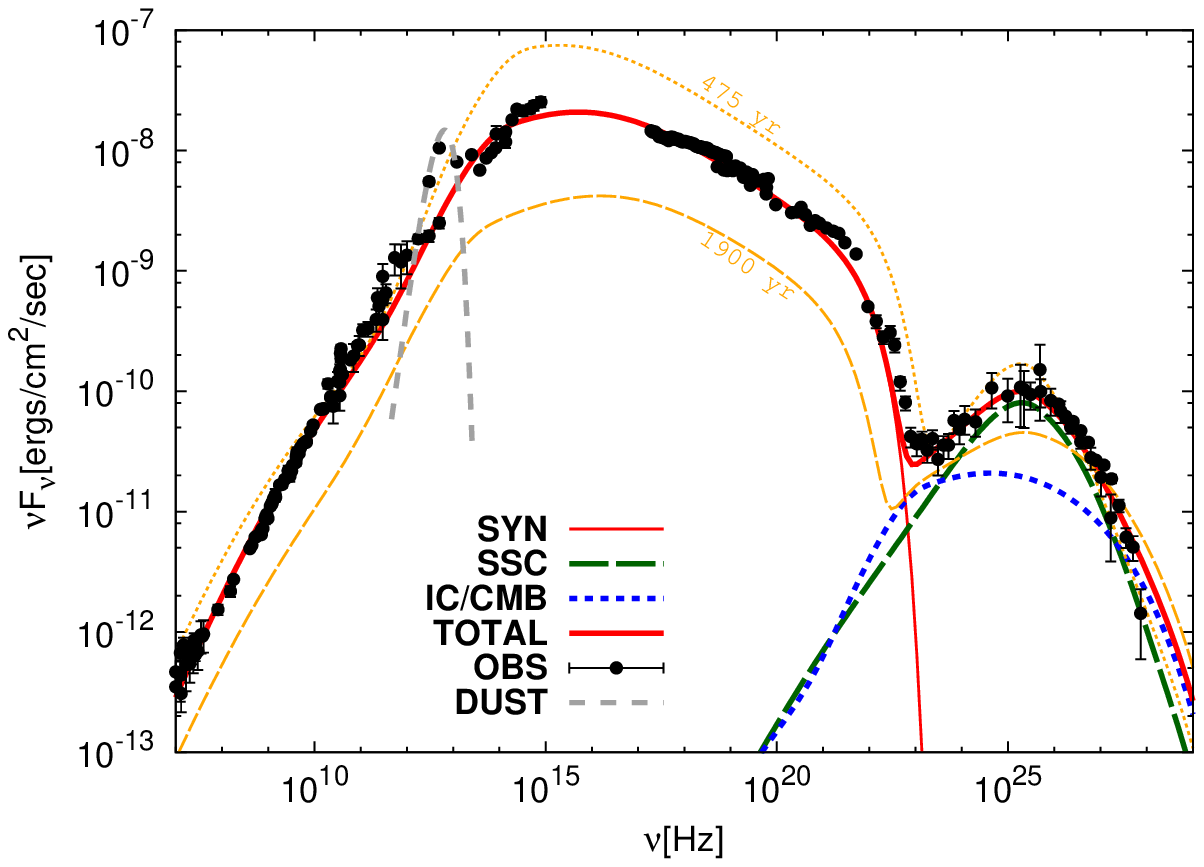}
\includegraphics[scale=0.6]{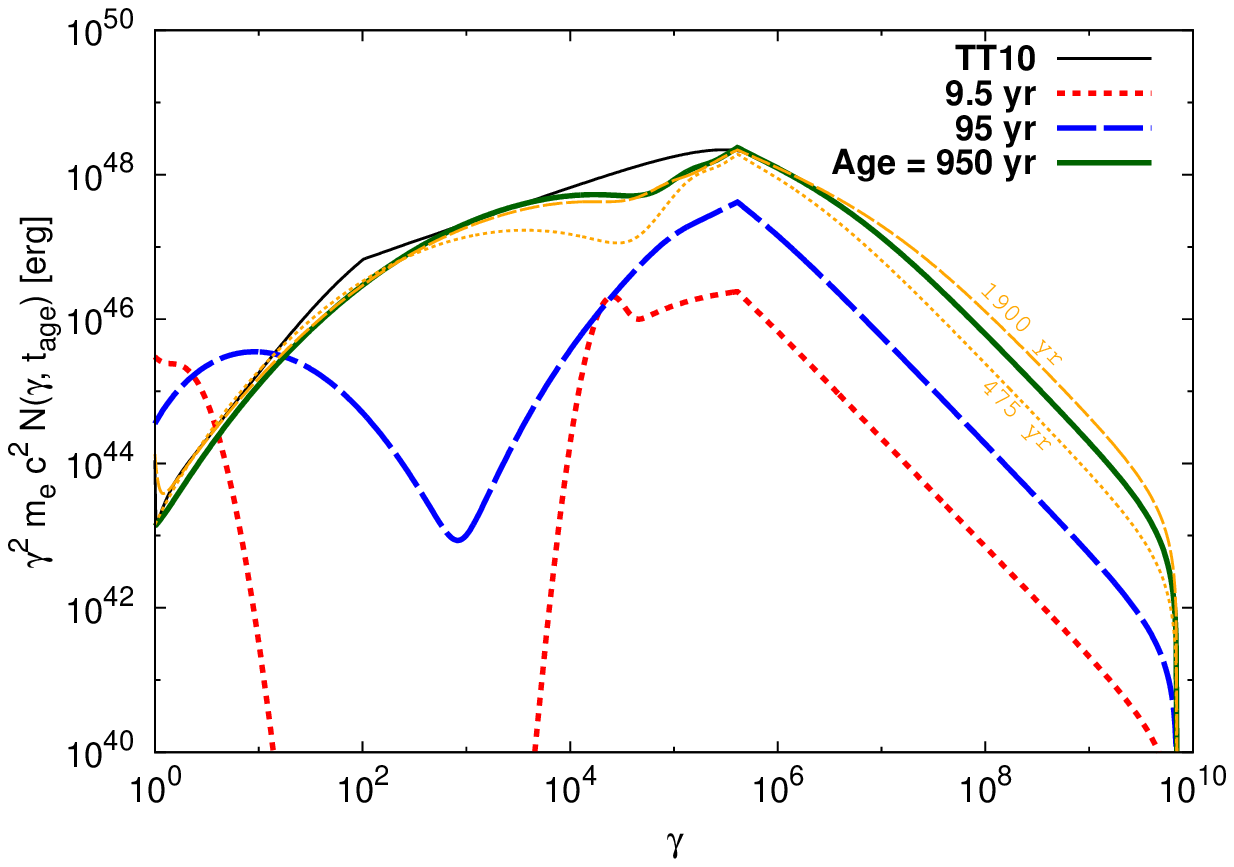} \\
\includegraphics[scale=0.6]{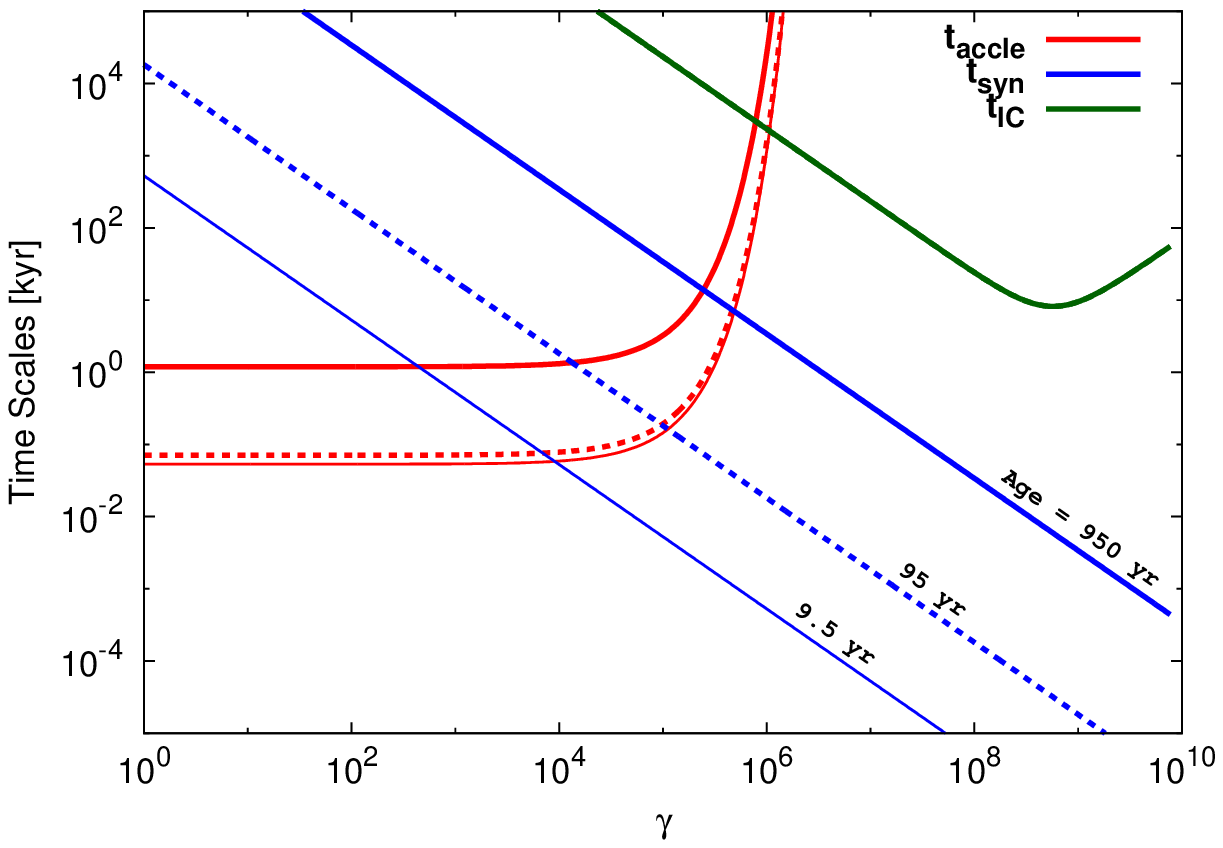}
\includegraphics[scale=0.6]{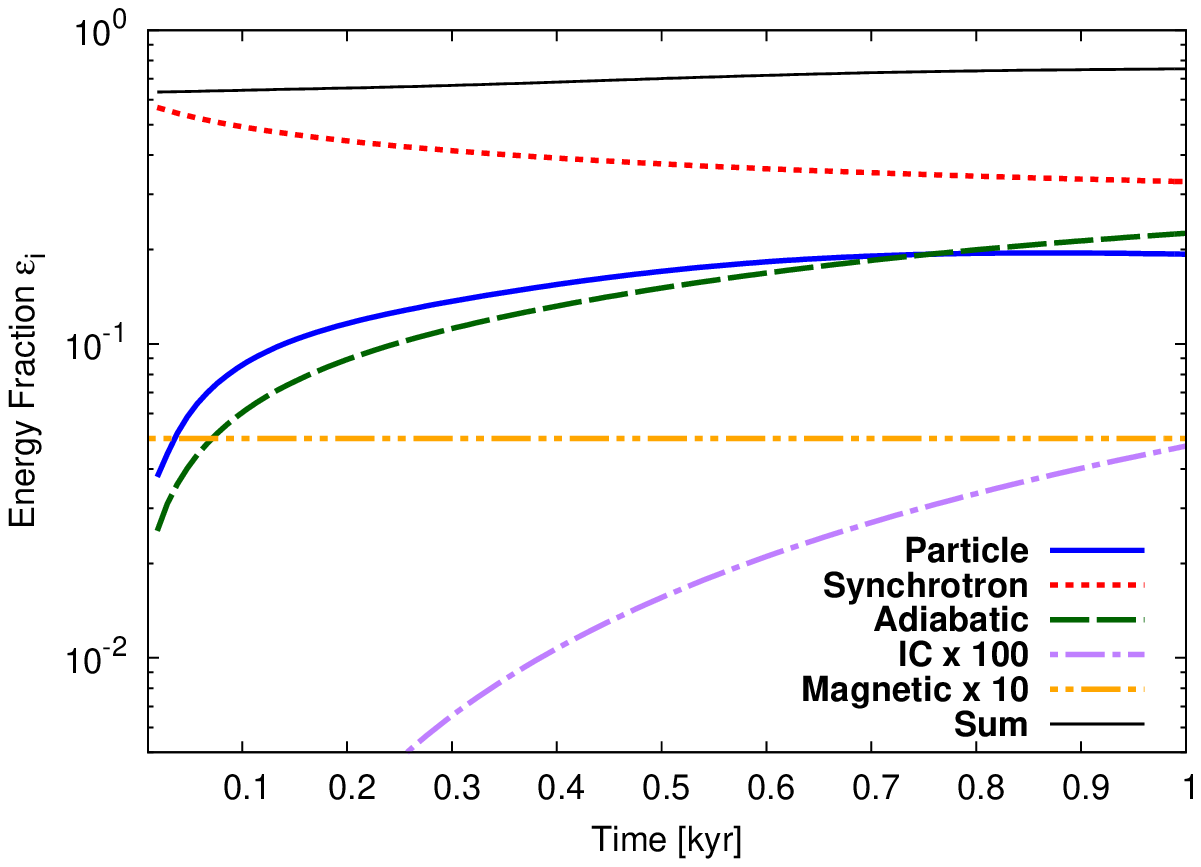}
\end{center}
\caption{
	Calculation results of Model 2: the current photon spectrum (top-left), evolution of the particle spectra (top-right), the time-scales (bottom-left), evolution of the fractional energy (bottom-right).
	See the caption in Figs. \ref{fig:SimpleDgg} and \ref{fig:model1} for details.
	Model 2 adopts $Q_{\rm ext} = Q_{\rm imp}$ and finite values of $\tau_{\rm turb}$ and $\gamma_{\rm cut}$ in Equation (\ref{eq:MomDiffCoefficient}).
	The used parameters are summarized in Table \ref{tbl:Parameters}.
\label{fig:model2}
}
\end{figure*}

Even for a model similar to the simple one in the previous section, a parameter choice of $\tau_{\rm turb} \sim \tau_{\rm acc,m} \sim t_{\rm age}$ may lead to agreement with both the observed spectrum and flux changes, but this accidental agreement of the time-scales seems unnatural.
Here, to explain the observed flux decreasing in radio, we consider models with a shorter duration of the stochastic acceleration than the age.
In order to agree with the observations including the decrease of the radio flux, a condition $\tau_{\rm acc,m} < \tau_{\rm turb} < t_{\rm age}$ is required irrespectively of $q$.
In this case, particles above $\gamma_{\rm min}$ will also be stochastically accelerated at $t=\tau_{\rm turb}$ and then the resultant becomes too hard in X-rays.
To suppress the acceleration of the X-ray emitting particles in our numerical code, we introduce the phenomenological parameter $\gamma_{\rm cut}$.
For example, the stochastic acceleration region is mostly on the rim of the PWN, while the X-ray emitting particles are located around the central part.
Note that, even for $q=5/3$, the required condition for the time-scales makes us introduce a finite value of $\gamma_{\rm cut}$.
Although we fix the index $q = 2$ in this section, we obtain similar results even for $q = 5/3$ with different values of other parameters (see Section \ref{sec:Summary}).

In this short duration model, the radio emission is considered to be the remnant of the past activity of the turbulence.
However, we cannot take an arbitrarily short duration $\tau_{\rm turb}$ for the Crab Nebula because of the energetics condition \citep[see discussion by][]{Kennel&Coroniti84b}.
If the radio-emitting particles were accelerated in very early phase of its evolution,
the required energy taking into account adiabatic cooling becomes larger than the energy deposited from the pulsar at $t=\tau_{\rm turb}\ll t_{\rm age}$.

As representative cases, we show the results for Models 1 and 2, whose parameters are in Table \ref{tbl:Parameters},
in Figures \ref{fig:model1} and \ref{fig:model2}, respectively.
The main difference in the two models is the injection of the extra particles;
the continuous injection $Q_{\rm ext} = Q_{\rm cont}$ for Model 1 and impulsive injection $Q_{\rm ext} = Q_{\rm imp}$ for Model 2.
For both of the models, the observed broadband spectrum (the top-left panels of Figures \ref{fig:model1} and \ref{fig:model2}) and the radio flux evolutions (Figure \ref{fig:FluxEvolution}) are well reproduced.
The energetics condition (see bottom-right panels of Figure \ref{fig:model1} and \ref{fig:model2}) is also satisfied, i.e., the sum of the energy fractions (thin black lines) is less than unity.

The parameters for $D_{\gamma \gamma}$ in Models 1 and 2 are almost common.
A few hundred years for $\tau_{\rm turb}$ and a few decades for $\tau_{\rm acc,m}$ are required.
While $\tau_{\rm acc,m}$ should be shorter than $\tau_{\rm turb}$, a too short $\tau_{\rm acc,m}$ leads to a too hard radio spectrum.
The combinations of $\tau_{\rm turb}$ and $\tau_{\rm acc,m}$ have been tuned to form the observed flat radio spectrum.
A choice $\gamma_{\rm cut} \sim \gamma_{\rm min}$ is required to accelerate only the radio-emitting particles, and then the parameters for $Q_{\rm PSR}$ are exactly the same for the both two models.
In order to compare the parameters of $Q_{\rm ext}$, we calculate the total particle number of the extra injection $N_{\rm ext} \equiv \int d\gamma dt Q_{\rm ext}(\gamma, t)$.
The values of $N_{\rm ext}$ and also $N_{\rm PSR} \equiv \int d \gamma d t Q_{\rm PSR}(\gamma, t)$ are tabulated in Table \ref{tbl:Parameters}.
The number $N_{\rm ext}$ in Model 1 is more than an order larger than those in the other models. This is because the extra particles are continuously injected even after the stop of the turbulence acceleration ($t>\tau_{\rm turb}$) in our model assumption.
As shown in the top-right panel of Figure \ref{fig:model1}, such lately injected particles pile up around $\gamma \sim \gamma_{\rm inj}$.
The existence of such low-energy particles does not affect the radio spectrum.

The current electron distribution in Model 1 (top-right panel of Figure \ref{fig:model1}) almost coincides with the result of TT10 at $\gamma > 10^2$, while that of Model 2 (top-right panel of Figure \ref{fig:model2}) is a smoothly curved spectrum for the radio-emitting particles like Figure \ref{fig:SimpleDgg}.
This difference arises from whether the extra injection is $Q_{\rm imp}(\gamma, t)$ or $Q_{\rm cont}(\gamma, t)$.
Although this apparent difference between the current electron distributions is almost dulled in the radiation spectra (top-left panels), the fitted result in Model 1 seems slightly better than Model 2.

\subsection{Allowed Parameter Range}\label{sec:Summary}

Both Models 1 and 2 successfully reproduce the observed broadband spectrum and the flux evolution within the required energetics.
However, parameter sets to reproduce the observations are not unique.
The tabulated parameters of $Q_{\rm ext}$ and $D_{\gamma \gamma}$ in Models 1 and 2 (adopting $q = 2$) are determined about 10 \% accuracy, i.e., the observed data allows 10 \% differences in the model parameters.
For the case of the continuous injection with $q = 5/3$, we also reproduce the observations with six times larger $\tau_{\rm acc,m}$ and an order of magnitude smaller $\dot{n}_{\rm cont}$ than those in Model 1, while the other parameters are the same as those in Model 1.

The general requirement of $Q_{\rm ext}$ is $\gamma_{\rm inj} \ll \gamma_{\rm min}$ and $N_{\rm ext} > 10^2 N_{\rm PSR}$ or equivalently $N_{\rm ext}$ is at least comparable to $N_{\rm PSR}$ in TT10.
Note that the value of $N_{\rm ext}$ is just a minimum number of the externally injected particles because not all the externally injected particles enter the stochastic acceleration process.
The combinations of $\gamma_{\rm inj}$ and $N_{\rm ext}$ are tuned to reproduce the radio flux.
The larger $\gamma_{\rm inj}$ leads to the smaller $N_{\rm ext}$.
The uncertainties of them are the same as the errors of the observed radio flux.

The observed decay of the radio flux is reproduced only by introducing $\tau_{\rm turb} < t_{\rm age}$.
Because the observed decay rate has relatively large error, a wide range of the time-scale, $\tau_{\rm turb} \sim 250 \pm 100$ yr, is acceptable.
On the other hand, $\tau_{\rm acc,m}$ or, precisely, the combination of $\tau_{\rm acc,m}$ and $\tau_{\rm turb}$ changes the radio spectral index.
This is similar to fitting the radio spectral index by tuning the power-law index $p_1$ for the low-energy component of the broken power-law particle spectrum at injection in TT10.
Because of the well-determined radio spectral slope of the Crab Nebula, $\tau_{\rm acc,m}$ should be tuned about 10 \% accuracy.
The smaller $\tau_{\rm acc,m}$ leads to a softer radio spectrum.

The emission below $\approx 10^{12}$ Hz is from the stochastically accelerated radio-emitting particles for all the results in Figures \ref{fig:SimpleDgg}, \ref{fig:model1} and \ref{fig:model2}.
We observe the spectral bump at infrared ($\approx 10^{12 - 13}$ Hz) in the spectrum of the Crab Nebula and the emission from this frequency range is dominated by the emission from the dust \citep[e.g.,][]{Owen&Barlow15}.
This apparent discontinuity of the broadband spectrum provides a room to consider the possible different origins for the radio-emitting and X-ray-emitting particles and allows to fit the broadband emission in our model.
On the other hand, the spectral data at ultraviolet ($\approx 10^{15 - 17}$ Hz) is lacked since the flux measurement of this frequency range is difficult as a result of the strong interstellar extinction.
This data gap also provides us the possibility that the emission below $\approx 10^{15}$ Hz is from the stochastically accelerated radio-emitting particles.
Although we tried fitting the emission below $\approx 10^{15}$ Hz by the stochastically accelerated particles and above $\approx 10^{17}$ Hz by $Q_{\rm PSR}$, we failed for any parameter sets of $Q_{\rm PSR}$, $Q_{\rm ext}$ and $D_{\gamma \gamma}$.
The smoothly curved high energy roll-off of the particle distribution by the stochastic acceleration inevitably overproduces the flux in X-rays of $\sim 10^{17-18}$ Hz.
In order to reproduce the observed spectral shape at the optical-ultraviolet regime of the Crab Nebula, we practically need two sharp breaks at $\gamma \sim 3 \times 10^5$ and $3 \times 10^6$.
The former and latter breaks correspond to the break energy of the broken power-law distribution $\gamma_{\rm b}$ in TT10 and the synchrotron cooling break, respectively.

\section{DISCUSSION}\label{sec:Discussion}



\subsection{Properties of Stochastic Acceleration Model}\label{sec:Properties}

The present stochastic acceleration model requires the tuning of the parameters of $Q_{\rm ext}$ and $D_{\gamma \gamma}$ for the energy of the radio-emitting particles to be comparable with that of the X-ray-emitting particles.
At present, we cannot judge whether the energy balance of the two components is accidental, or some unveiled tuning mechanism has been working.
However, we emphasize that, even when the two particle components are separated, the photon spectrum does not necessarily show a distinctive two components.
As shown in the green lines of the top-right panel of Figure \ref{fig:SimpleDgg}, we do not need to tune the parameters to make the two particle components merge perfectly as a broken power-law distribution.

Once the parameters are tuned, the continuity of the broadband spectra produced by the two distinct components becomes persistent.
The yellow lines of the photon spectra (top-left panels) in Figures \ref{fig:model1} and \ref{fig:model2} are the spectra at a half and twice of $t_{\rm age}$, respectively.
The corresponding particle spectra are also found on the top-right panels in Figures \ref{fig:model1} and \ref{fig:model2}.
The stochastic acceleration lasts until $\sim \tau_{\rm turb} \sim$ a few hundreds years, while the injection from the pulsar $Q_{\rm PSR}$ is effective until $\sim \tau_0 \sim$ 700 yr.
Because they last the comparable duration, the broadband spectrum becomes apparently continuous beyond $t > \tau_{\rm turb}$.
We find that the radio-emitting and X-ray-emitting particles have a comparable energy even at $10 t_{\rm age}$.

We find that the stochastic acceleration reproduces the flat radio spectrum with the smoothly curved spectrum, which is sensitive to relevant time-scales (bottom-left panels of Figures \ref{fig:model1} and \ref{fig:model2}).
Although the flat radio spectrum is the universal feature in PWNe, its spectral indices are different between objects.
In other words, slightly different sets of $\tau_{\rm acc,m}$ and $\tau_{\rm turb}$ are enough to explain the diversity of the observed radio spectral index between different PWNe.
\subsection{Obtained Parameter Values of Radio-Emitting Particles}\label{sec:Values}

In our models, the acceleration time-scales at $\gamma=\gamma_{\rm min}$ are 25--950 yr, which are much longer than that required in the stochastic acceleration models in blazars, where the acceleration time-scale ($10^4$--$10^5$ s in 3C 279) is comparable to the dynamical time-scale \citep[e.g.][]{Asano&Hayashida15}.
On the other hand, to explain the Fermi Bubbles by the leptonic model with the stochastic acceleration, the required acceleration time-scale ($0.1$--$1$ Myr) is less than the dynamical time-scale $\sim 10$ Myr \citep[][]{Sasaki+15}.
The relatively long acceleration time-scale required in this paper does not seem difficult to be realized.
In the stochastic acceleration via turbulence, the acceleration time-scale depends on the typical scale of the turbulence.
Below, we discuss the acceleration time-scale and the turbulence property to confirm that the model assumption is not extreme.

In Equation (\ref{eq:MomDiffCoefficient}), we have assumed that the diffusion coefficient $D_{\gamma \gamma}$ is almost constant for $t<\tau_{\rm turb}$.
Since the nebula is expanding with time, the assumption of the constant $D_{\gamma \gamma}$ is not necessarily trivial.
Let us assume that the turbulence is injected at a smaller scale than $R_{\rm PWN}(t)=v_{\rm PWN} t$; here we take the minimum wave number of the turbulence as $k_{\rm min}=1/(\epsilon R_{\rm PWN})$ with a dimensionless factor $\epsilon < 1$ .
The turbulence velocity at injection may be mildly relativistic $v_{\rm turb}\sim c / \sqrt{3}$ \citep{Porth+16}.
The injected energy may cascade to higher wave numbers following the Kolmogorov spectrum.
In this case, given the Alfv\'en velocity $v_{\rm A}$, the energy density at $k=k_0 \equiv k_{\rm min} (v_{\rm turb}/v_{\rm A})^3$ becomes comparable to the energy density of the background magnetic field \citep[e.g.,][]{Cho&Vishniac00, Inoue+11}.
At this scale, the transition from the hydrodynamic Kolmogorov
cascade to the magnetohydrodynamic critical balance cascade occurs \citep{Goldreich&Sridhar95}.
This transition induces the steepening of the power spectrum of the turbulence at $k=k_0$.
The energy diffusion of particles would be dominated by the interaction with the turbulence at this scale $l_0=k_0^{-1}$ \citep[c.f.,][]{Dupree66, Lynn+14}.
As will be shown, $l_0$ would be larger than the Larmor radius of particles of $\gamma_{\rm min}$, which means the non-resonant
interaction, i.e., supporting the hard sphere assumption.
The effective velocity of the turbulence at scale $l_0$
is $\sim v_{\rm A}$, from which we obtain the energy gain
factor per scattering as $\Delta \gamma/\gamma \sim (v_{\rm A}/c)^2$
and interaction time-scale $\Delta t \sim l_0/c$.
The above simple model implies the energy-independent acceleration time-scale as
\begin{eqnarray}\label{eq:Accmodel0}
	\tau_{\rm acc}
	& \sim &
	2 \frac{l_0}{c} \left( \frac{c}{v_{\rm A}} \right)^2.
\end{eqnarray}
The Alfv\'en velocity in the turbulence region depends on the proton density.
In the continuous injection model, protons from the SN ejecta are also injected.
Introducing the number fraction $f_{\rm e}$ of the stochastically accelerated particles,
the proton density at the edge of the nebula is written as $n_{\rm p}=f_{\rm e}^{-1} \dot{n}_{\rm cont} (t/t_{\rm age})^2/(4 \pi R_{\rm PWN}^2 v_{\rm PWN})=\dot{n}_{\rm cont}/(4 \pi f_{\rm e} t_{\rm age}^2 v_{\rm PWN}^3)$.
For $t \ll \tau_{\rm turb}$, the magnetic field is estimated as $B(t)=\sqrt{6 \eta_B L_0 t/R_{\rm PWN}^3}$, so that
\begin{eqnarray}\label{eq:alv}
	v_{\rm A}^2=\frac{B^2}{4 \pi m_{\rm p} n_{\rm p}}=\frac{6 \eta_B f_{\rm e} L_0 t_{\rm age}^2}{m_{\rm p} \dot{n}_{\rm cont} t^2}.
\end{eqnarray}
Then, the equation (\ref{eq:Accmodel0}) provides
\begin{eqnarray}\label{eq:Accmodel}
	\tau_{\rm acc}
	& \sim &
	\frac{2 \epsilon c v_{\rm A} v_{\rm PWN} t}{v_{\rm turb}^3} \\
	&\simeq&
31~\mbox{yr} \left( \frac{\epsilon}{0.1} \right) \left( \frac{\eta_B f_{\rm e}}{5 \times 10^{-6}} \right)^{1/2} \left( \frac{L_0}{3.4 \times 10^{39}~\mbox{erg}~\mbox{s}^{-1}} \right)^{1/2} \nonumber \\
&&\times \left( \frac{t_{\rm age}}{950~\mbox{yr}} \right) \left( \frac{\dot{n}_{\rm cont}}{6 \times 10^{42}~\mbox{s}^{-1}} \right)^{-1/2} \left( \frac{v_{\rm turb}}{0.05c} \right)^{-3},
\end{eqnarray}
which is constant in time. The above value indicates that the combination $f_{\rm e} \sim 10^{-3}$ and $\epsilon \sim 10^{-1}$ is preferable to agree with the acceleration time-scale $\tau_{\rm acc,m}$ for the model parameters in Models 1 and 2.
For this parameter set, the scale
\begin{eqnarray}\label{eq:l0model}
	l_0
	& \sim &
	1.5 \times 10^{16} ~\mbox{cm} \left( \frac{\epsilon}{0.1} \right) \left( \frac{\eta_B f_{\rm e}}{5 \times 10^{-6}} \right)^{3/2} \nonumber \\
	&&\times \left( \frac{L_0}{3.4 \times 10^{39}~\mbox{erg}~\mbox{s}^{-1}} \right)^{3/2} \left( \frac{t_{\rm age}}{950~\mbox{yr}} \right)^3 \nonumber \\
	&&\times \left( \frac{\dot{n}_{\rm cont}}{6 \times 10^{42}~\mbox{s}^{-1}} \right)^{-3/2} \left( \frac{v_{\rm turb}}{0.05c} \right)^{-3} \left( \frac{t}{100~\mbox{yr}} \right)^{-2},
\end{eqnarray}
is longer than the Larmor radius of particles $\sim 5 \times 10^{11}$ cm for $\gamma=\gamma_{\rm min}$ at $t=100$ yr.
This large-scale turbulence is consistent with the hard sphere assumption,
while the resonant interaction seems difficult in this case.
The above rough estimate may be rather simplistic, but shows that the parameter sets in our models are not in impractical ranges.

As for the extra particle injection, the injection rates required in the continuous injection models ($Q_{\rm ext}=Q_{\rm cont}$) are reasonable as well.
As obtained above, the densities of the extra particles in the continuous acceleration model and Model 1 are $n_{\rm p} \sim 10^{-3}$--$10^{-1} (f_{\rm e}/10^{-3})^{-1}~\mbox{cm}^{-3}$,
which is much less than the ejecta density (typically $n_{\rm ej} \sim 1~{\rm cm^{-3}}$).
The photoionization process of the neutral particles penetrating into the PWN may provide such a density.
A small fraction $f_{\rm e}$ of those particles may be mildly relativistic and injected into the acceleration process.
For the impulsive injection model, the required number $N_{\rm ext}$ in Model 2 seems not easy supply.
Let us estimate the maximum value for $N_{\rm ext}$ assuming that a fraction of the initial rotational energy is converted into the pairs by two-photon annihilation process \citep[e.g.,][]{Metzger+14,Murase+15}.
We set the typical (bulk) Lorentz factor of the particles is $\Gamma_{\rm b} \approx 10^6$ \citep[c.f.,][]{Kennel&Coroniti84b}.
Considering the very early phase of evolution, we assume the equipartition between the particles and magnetic field.
This is the fairly optimistic condition for the efficient pair production because the spin-down power is the most effectively converted to the synchrotron photons for $\eta_{\rm e} \approx \eta_{\rm B} \approx 0.5$.
In the very early stage, the SN ejecta expands with a constant velocity $v_{\rm ej}\sim 5 \times 10^3~\mbox{km}~\mbox{s}^{-1}$.
The nebula size may be comparable to the radius of the SN ejecta $R_{\rm ej}=v_{\rm ej} t$,
then the magnetic field in the nebula is $B\sim \sqrt{3 L_0 t/R_{\rm ej}^3}\simeq 0.1~{\rm G} (v_{\rm ej}/5 \times 10^3~\mbox{km}~\mbox{s}^{-1})^{-3/2} (t/\mbox{yr})^{-1}$.
The typical synchrotron energy $\varepsilon_0 = \Gamma^2_{\rm b} \hbar e B/(m_{\rm e} c)$ can be larger than $m_{\rm e} c^2$ for
\begin{eqnarray}\label{eq:esyn}
	t \leq t_{\gamma \gamma} \equiv 6.5 \times 10^{-3}~\mbox{yr} \left( \frac{\Gamma_{\rm b}}{10^6} \right)^{2}
\left( \frac{v_{\rm ej}}{5 \times 10^3 \mbox{km}~\mbox{s}^{-1}} \right)^{-3/2},
\end{eqnarray}
at which the released energy as particles, $L_0 t_{\gamma \gamma}/2 \simeq 3.5 \times 10^{44}$ erg, is comparable to the model requirement $N_{\rm ext} m_{\rm e} c^2 \approx 1.6 \times 10^{45}$ erg.
The synchrotron cooling time $6 \pi m_{\rm e} c/(\sigma_{\rm T} B^2 \Gamma_{\rm b})$ is much shorter than the age $t < t_{\gamma \gamma}$, so that almost all the released energy
is converted into synchrotron photons.
However, even if those photons with energies $\geq m_{\rm e} c^2$ are confined in the nebula,
the photon density $n_\gamma(t_{\gamma \gamma})=3 L_0 t_{\gamma \gamma}/(m_{\rm e} c^2)/(8 \pi R_{\rm ej}^3)$ is so low
that the optical depth for $\gamma \gamma$-absorption $\sim 0.1 \sigma_{\rm T} n_\gamma(t_{\gamma \gamma}) R_{\rm ej}$
is much less than unity as $\simeq 6 \times 10^{-4} (\Gamma_{\rm b}/10^6)^{-2} (v_{\rm ej}/5 \times 10^3 \mbox{km}~\mbox{s}^{-1})^{-1/2}$.
We cannot expect efficient pair production for $t>t_{\gamma \gamma}$, because the optical depth decreases as $\propto t^{-1}$,
and the typical photon energy also decays below $m_{\rm e} c^2$.
A smaller $\Gamma_{\rm b}$ does not resolve the problem because, in this case, a shorter $t_{\gamma \gamma}$ does not satisfy the necessary requirement $L_0 t_{\gamma \gamma}/2>N_{\rm ext} m_{\rm e} c^2$.
An extreme model in the very early phase may be required to inject particles of $N_{\rm ext}\sim 10^{51}$ impulsively.
For example, the millisecond magnetar-like behavior of the Crab Pulsar, which does not affect the present broadband spectrum \citep[c.f.,][]{Tanaka16}.

Although the above discussion suggests that Model 1 (hard sphere, continuous injection, and short duration) is the most reasonable one, we should note caveats below.
While most of PWNe have their radio luminosities comparable to their X-ray ones, in the present model of two different origins of particles, the mechanism to much the luminosities is not resolved.
The radio luminosity is exactly proportional to the particle number supply ($N_{\rm imp}$ and $\dot{n}_{\rm cont}$).
In addition to the number supply, the energy supply will be considered properly in the future studies.
In the present model, the duration of the acceleration has been treated as an unknown parameter in our calculation.
We have required the turbulence to decay after the appropriate amount of the energy is transferred to the radio-emitting particles.
The feedback from the accelerated particles may decay the turbulence, or the increase of inertia due to the proton injection ($\propto t^2$ in Model 1) may prevent excitation of significantly fast turbulence with $v_{\rm turb} \sim 0.05 c$ in the later phase.

The actual PWNe may be inhomogeneous, and the turbulence acceleration region may be unevenly distributed.
To simulate such situations, more complex models such as two-zone or 1-D models may be desired.
The back reaction to the turbulence may be an important issue to be included in calculation.
At present, however, the increase of the parameters required in such models seems premature,
though such models may not need some phenomenological parameters such as $\gamma_{\rm cut}$ or $\tau_{\rm turb}$.
Future observations or multi-dimensional MHD simulations, which can resolve significantly small eddy, will be demanded to develop the model further.

\section{Conclusions}\label{sec:Conclusions}

We examine the stochastic acceleration model of the radio-emitting particles in the Crab Nebula.
The model and method in this paper are simple extension of the one-zone time-dependent model in TT10.
For the X-ray-emitting particles, we attribute the particles accelerated at the pulsar wind termination shock $Q_{\rm PSR}$  according to \citet{Kennel&Coroniti84b}.
On the other hand, for the radio-emitting particles, we introduce the extra particle injection $Q_{\rm ext}$ and the energy diffusion $D_{\gamma \gamma}$ due to turbulence.
The synchrotron radiation from the two distinct components is adjusted to connect smoothly at the infrared bump ($\approx 10^{12-13}$ Hz) and not at the ultraviolet data gap ($\approx 10^{15-17}$ Hz).
The stochastic acceleration model can resolve the two long-standing problems of PWNe, i.e., the too many number of the particles supplied from the central pulsar and the formation of the observed flat radio spectrum \citep[c.f.,][]{Kennel&Coroniti84b}.

Our results demonstrate that the stochastic acceleration process can reproduce the observed spectrum for certain sets of the parameters of $Q_{\rm ext}$ and $D_{\gamma \gamma}$.
Although the obtained values of the model parameters are just a few of the examples and not strictly constrained, our results require conditions for the stochastic acceleration as following.
The normalization $N_{\rm imp}$ or $\dot{n}_{\rm cont}$ (Equation (\ref{eq:SNinjection})) should be tuned to reproduce the observed radio flux, while the time-scales $\tau_{\rm acc,m}$ and $\tau_{\rm turb}$ (Equation (\ref{eq:MomDiffCoefficient})) are tuned to reproduce the observed radio spectral index and the observed radio flux decay.
The total particle number of the external injection $N_{\rm ext}$ is at least two orders of magnitude larger than the particles supplied from the pulsar $N_{\rm PSR}$.
The shorter duration $\tau_{\rm turb}$ demands the faster energy injection to the radio-emitting particles by a shorter $\tau_{\rm acc, m}$, and then an too small $\tau_{\rm turb}$ violates the energy budget restricted by the released energy from the central pulsar.
The effective duration of the turbulence acceleration should be longer than a few hundred years.
The acceleration time-scale should be shorter than $\tau_{\rm turb}$, while the diffusion index $q$ is not well constrained.

The spectral component due to the stochastically accelerated particles is required to smoothly match the component from the pulsar wind.
This seems accidental, and the parameter tuning is required in the present model.
However, some self-regulated feedback mechanism to adjust the energy budgets of the two components may exist.
A simulation including the back reaction to the turbulence will be required as a future study.

\section*{Acknowledgments}

S. J. T. would like to thank Y. Ohira and K. Murase for useful discussion.
This work is supported by Grants-in-Aid for Scientific Research Nos.
15H05440 (ST), 15K05069 and 16K05291 (KA).

\bibliography{draft}

\begin{thebibliography}{}
\expandafter\ifx\csname natexlab\endcsname\relax\def\natexlab#1{#1}\fi

\bibitem[{{Abdo} {et~al.}(2010){Abdo}, {Ackermann}, {Ajello}, {Atwood},
  {Axelsson}, {Baldini}, {Ballet}, {Barbiellini}, {Baring}, {Bastieri},
  {Bechtol}, {Bellazzini}, {Berenji}, {Blandford}, {Bloom}, {Bonamente},
  {Borgland}, {Bregeon}, {Brez}, {Brigida}, {Bruel}, {Burnett}, {Caliandro},
  {Cameron}, {Camilo}, {Caraveo}, {Casandjian}, {Cecchi}, {{\c C}elik},
  {Chekhtman}, {Cheung}, {Chiang}, {Ciprini}, {Claus}, {Cognard},
  {Cohen-Tanugi}, {Cominsky}, {Conrad}, {Dermer}, {de Angelis}, {de Luca}, {de
  Palma}, {Digel}, {Silva}, {Drell}, {Dubois}, {Dumora}, {Espinoza}, {Farnier},
  {Favuzzi}, {Fegan}, {Ferrara}, {Focke}, {Frailis}, {Freire}, {Fukazawa},
  {Funk}, {Fusco}, {Gargano}, {Gasparrini}, {Gehrels}, {Germani}, {Giavitto},
  {Giebels}, {Giglietto}, {Giordano}, {Glanzman}, {Godfrey}, {Grenier},
  {Grondin}, {Grove}, {Guillemot}, {Guiriec}, {Hanabata}, {Harding},
  {Hayashida}, {Hays}, {Hughes}, {J{\'o}hannesson}, {Johnson}, {Johnson},
  {Johnson}, {Johnson}, {Johnston}, {Kamae}, {Katagiri}, {Kataoka}, {Kawai},
  {Kerr}, {Kn{\"o}dlseder}, {Kocian}, {Kramer}, {Kuehn}, {Kuss}, {Lande},
  {Latronico}, {Lee}, {Lemoine-Goumard}, {Longo}, {Loparco}, {Lott},
  {Lovellette}, {Lubrano}, {Lyne}, {Makeev}, {Marelli}, {Mazziotta}, {McEnery},
  {Meurer}, {Michelson}, {Mitthumsiri}, {Mizuno}, {Moiseev}, {Monte},
  {Monzani}, {Moretti}, {Morselli}, {Moskalenko}, {Murgia}, {Nakamori},
  {Nolan}, {Norris}, {Noutsos}, {Nuss}, {Ohsugi}, {Omodei}, {Orlando}, {Ormes},
  {Ozaki}, {Paneque}, {Panetta}, {Parent}, {Pelassa}, {Pepe}, {Pesce-Rollins},
  {Pierbattista}, {Piron}, {Porter}, {Rain{\`o}}, {Rando}, {Ray}, {Razzano},
  {Reimer}, {Reimer}, {Reposeur}, {Ritz}, {Rochester}, {Rodriguez}, {Romani},
  {Roth}, {Ryde}, {Sadrozinski}, {Sanchez}, {Sander}, {Saz Parkinson},
  {Scargle}, {Sgr{\`o}}, {Siskind}, {Smith}, {Smith}, {Spandre}, {Spinelli},
  {Stappers}, {Strickman}, {Suson}, {Tajima}, {Takahashi}, {Tanaka}, {Thayer},
  {Thayer}, {Theureau}, {Thompson}, {Thorsett}, {Tibaldo}, {Torres}, {Tosti},
  {Tramacere}, {Uchiyama}, {Usher}, {Van Etten}, {Vasileiou}, {Vilchez},
  {Vitale}, {Waite}, {Wallace}, {Wang}, {Watters}, {Weltevrede}, {Winer},
  {Wood}, {Ylinen}, \& {Ziegler}}]{Abdo+10}
{Abdo}, A.~A., {Ackermann}, M., {Ajello}, M., {et~al.} 2010, \apj, 708, 1254

\bibitem[{{Aharonian} {et~al.}(2006){Aharonian}, {Akhperjanian}, {Bazer-Bachi},
  {Beilicke}, {Benbow}, {Berge}, {Bernl{\"o}hr}, {Boisson}, {Bolz}, {Borrel},
  {Braun}, {Breitling}, {Brown}, {B{\"u}hler}, {B{\"u}sching}, {Carrigan},
  {Chadwick}, {Chounet}, {Cornils}, {Costamante}, {Degrange}, {Dickinson},
  {Djannati-Ata{\"\i}}, {O'C.~Drury}, {Dubus}, {Egberts}, {Emmanoulopoulos},
  {Espigat}, {Feinstein}, {Ferrero}, {Fiasson}, {Fontaine}, {Funk}, {Funk},
  {Gallant}, {Giebels}, {Glicenstein}, {Goret}, {Hadjichristidis}, {Hauser},
  {Hauser}, {Heinzelmann}, {Henri}, {Hermann}, {Hinton}, {Hofmann}, {Holleran},
  {Horns}, {Jacholkowska}, {de Jager}, {Kh{\'e}lifi}, {Komin}, {Konopelko},
  {Kosack}, {Latham}, {Le Gallou}, {Lemi{\`e}re}, {Lemoine-Goumard}, {Lohse},
  {Martin}, {Martineau-Huynh}, {Marcowith}, {Masterson}, {McComb}, {de
  Naurois}, {Nedbal}, {Nolan}, {Noutsos}, {Orford}, {Osborne}, {Ouchrif},
  {Panter}, {Pelletier}, {Pita}, {P{\"u}hlhofer}, {Punch}, {Raubenheimer},
  {Raue}, {Rayner}, {Reimer}, {Reimer}, {Ripken}, {Rob}, {Rolland}, {Rowell},
  {Sahakian}, {Saug{\'e}}, {Schlenker}, {Schlickeiser}, {Schwanke}, {Sol},
  {Spangler}, {Spanier}, {Steenkamp}, {Stegmann}, {Superina}, {Tavernet},
  {Terrier}, {Th{\'e}oret}, {Tluczykont}, {van Eldik}, {Vasileiadis}, {Venter},
  {Vincent}, {V{\"o}lk}, {Wagner}, \& {Ward}}]{Aharonian+06}
{Aharonian}, F., {Akhperjanian}, A.~G., {Bazer-Bachi}, A.~R., {et~al.} 2006,
  \aap, 457, 899

\bibitem[{{Albert} {et~al.}(2008){Albert}, {Aliu}, {Anderhub}, {Antoranz},
  {Armada}, {Baixeras}, {Barrio}, {Bartko}, {Bastieri}, {Becker}, {Bednarek},
  {Berger}, {Bigongiari}, {Biland}, {Bock}, {Bordas}, {Bosch-Ramon}, {Bretz},
  {Britvitch}, {Camara}, {Carmona}, {Chilingarian}, {Coarasa}, {Commichau},
  {Contreras}, {Cortina}, {Costado}, {Curtef}, {Danielyan}, {Dazzi}, {De
  Angelis}, {Delgado}, {de los Reyes}, {De Lotto}, {Domingo-Santamar{\'{\i}}a},
  {Dorner}, {Doro}, {Errando}, {Fagiolini}, {Ferenc}, {Fern{\'a}ndez}, {Firpo},
  {Flix}, {Fonseca}, {Font}, {Fuchs}, {Galante}, {Garc{\'{\i}}a-L{\'o}pez},
  {Garczarczyk}, {Gaug}, {Giller}, {Goebel}, {Hakobyan}, {Hayashida},
  {Hengstebeck}, {Herrero}, {H{\"o}hne}, {Hose}, {Hsu}, {Jacon}, {Jogler},
  {Kosyra}, {Kranich}, {Kritzer}, {Laille}, {Lindfors}, {Lombardi}, {Longo},
  {L{\'o}pez}, {L{\'o}pez}, {Lorenz}, {Majumdar}, {Maneva}, {Mannheim},
  {Mansutti}, {Mariotti}, {Mart{\'{\i}}nez}, {Mazin}, {Merck}, {Meucci},
  {Meyer}, {Miranda}, {Mirzoyan}, {Mizobuchi}, {Moralejo}, {Nieto}, {Nilsson},
  {Ninkovic}, {O{\~n}a-Wilhelmi}, {Otte}, {Oya}, {Paneque}, {Panniello},
  {Paoletti}, {Paredes}, {Pasanen}, {Pascoli}, {Pauss}, {Pegna}, {Persic},
  {Peruzzo}, {Piccioli}, {Poller}, {Prandini}, {Puchades}, {Raymers}, {Rhode},
  {Rib{\'o}}, {Rico}, {Rissi}, {Robert}, {R{\"u}gamer}, {Saggion},
  {S{\'a}nchez}, {Sartori}, {Scalzotto}, {Scapin}, {Schmitt}, {Schweizer},
  {Shayduk}, {Shinozaki}, {Shore}, {Sidro}, {Sillanp{\"a}{\"a}}, {Sobczynska},
  {Stamerra}, {Stark}, {Takalo}, {Temnikov}, {Tescaro}, {Teshima}, {Tonello},
  {Torres}, {Turini}, {Vankov}, {Vitale}, {Wagner}, {Wibig}, {Wittek},
  {Zandanel}, {Zanin}, \& {Zapatero}}]{Albert+08}
{Albert}, J., {Aliu}, E., {Anderhub}, H., {et~al.} 2008, \apj, 674, 1037

\bibitem[{{Arons}(2012)}]{Arons12}
{Arons}, J. 2012, \ssr, 173, 341

\bibitem[{{Asano} \& {Hayashida}(2015)}]{Asano&Hayashida15}
{Asano}, K., \& {Hayashida}, M. 2015, \apjl, 808, L18

\bibitem[{{Asano} {et~al.}(2014){Asano}, {Takahara}, {Kusunose}, {Toma}, \&
  {Kakuwa}}]{Asano+14}
{Asano}, K., {Takahara}, F., {Kusunose}, M., {Toma}, K., \& {Kakuwa}, J. 2014,
  \apj, 780, 64

\bibitem[{{Atoyan} \& {Aharonian}(1996)}]{Atoyan&Aharonian96}
{Atoyan}, A.~M., \& {Aharonian}, F.~A. 1996, \mnras, 278, 525

\bibitem[{{Aumont} {et~al.}(2010){Aumont}, {Conversi}, {Thum}, {Wiesemeyer},
  {Falgarone}, {Mac{\'{\i}}as-P{\'e}rez}, {Piacentini}, {Pointecouteau},
  {Ponthieu}, {Puget}, {Rosset}, {Tauber}, \& {Tristram}}]{Aumont+10}
{Aumont}, J., {Conversi}, L., {Thum}, C., {et~al.} 2010, \aap, 514, A70

\bibitem[{{Baars} {et~al.}(1977){Baars}, {Genzel}, {Pauliny-Toth}, \&
  {Witzel}}]{Baars+77}
{Baars}, J.~W.~M., {Genzel}, R., {Pauliny-Toth}, I.~I.~K., \& {Witzel}, A.
  1977, \aap, 61, 99

\bibitem[{{Baldwin}(1971)}]{Baldwin71}
{Baldwin}, J.~E. 1971, in IAU Symposium, Vol.~46, The Crab Nebula, ed. R.~D.
  {Davies} \& F.~{Graham-Smith}, 22

\bibitem[{{Bamba} {et~al.}(2010){Bamba}, {Anada}, {Dotani}, {Mori}, {Yamazaki},
  {Ebisawa}, \& {Vink}}]{Bamba+10}
{Bamba}, A., {Anada}, T., {Dotani}, T., {et~al.} 2010, \apjl, 719, L116

\bibitem[{{Begelman}(1998)}]{Begelman98}
{Begelman}, M.~C. 1998, \apj, 493, 291

\bibitem[{{Bietenholz} \& {Kronberg}(1991)}]{Bietenholz&Kronberg91}
{Bietenholz}, M.~F., \& {Kronberg}, P.~P. 1991, \apj, 368, 231

\bibitem[{{Blandford} \& {Eichler}(1987)}]{Blandford&Eichler87}
{Blandford}, R., \& {Eichler}, D. 1987, \physrep, 154, 1

\bibitem[{{Blondin} {et~al.}(2001){Blondin}, {Chevalier}, \&
  {Frierson}}]{Blondin+01}
{Blondin}, J.~M., {Chevalier}, R.~A., \& {Frierson}, D.~M. 2001, \apj, 563, 806

\bibitem[{{Bucciantini} {et~al.}(2011){Bucciantini}, {Arons}, \&
  {Amato}}]{Bucciantini+11}
{Bucciantini}, N., {Arons}, J., \& {Amato}, E. 2011, \mnras, 410, 381

\bibitem[{{Bucciantini} {et~al.}(2005){Bucciantini}, {del Zanna}, {Amato}, \&
  {Volpi}}]{Bucciantini+05}
{Bucciantini}, N., {del Zanna}, L., {Amato}, E., \& {Volpi}, D. 2005, \aap,
  443, 519

\bibitem[{{Cho} \& {Vishniac}(2000)}]{Cho&Vishniac00}
{Cho}, J., \& {Vishniac}, E.~T. 2000, \apj, 538, 217

\bibitem[{{Daugherty} \& {Harding}(1982)}]{Daugherty&Harding82}
{Daugherty}, J.~K., \& {Harding}, A.~K. 1982, \apj, 252, 337

\bibitem[{{de Jager}(2008)}]{deJager08}
{de Jager}, O.~C. 2008, \apjl, 678, L113

\bibitem[{{de Jager} {et~al.}(2008){de Jager}, {Slane}, \&
  {LaMassa}}]{deJager+08}
{de Jager}, O.~C., {Slane}, P.~O., \& {LaMassa}, S. 2008, \apjl, 689, L125

\bibitem[{{Del Zanna} {et~al.}(2004){Del Zanna}, {Amato}, \&
  {Bucciantini}}]{DelZanna+04}
{Del Zanna}, L., {Amato}, E., \& {Bucciantini}, N. 2004, \aap, 421, 1063

\bibitem[{{Dupree}(1966)}]{Dupree66}
{Dupree}, T.~H. 1966, Physics of Fluids, 9, 1773

\bibitem[{{Fang} \& {Zhang}(2010)}]{Fang&Zhang10}
{Fang}, J., \& {Zhang}, L. 2010, \aap, 515, A20

\bibitem[{{Fermi}(1949)}]{Fermi49}
{Fermi}, E. 1949, Physical Review, 75, 1169

\bibitem[{{Fleishman} \& {Bietenholz}(2007)}]{Fleishman&Bietenholz07}
{Fleishman}, G.~D., \& {Bietenholz}, M.~F. 2007, \mnras, 376, 625

\bibitem[{{Gelfand} {et~al.}(2015){Gelfand}, {Slane}, \& {Temim}}]{Gelfand+15}
{Gelfand}, J.~D., {Slane}, P.~O., \& {Temim}, T. 2015, \apj, 807, 30

\bibitem[{{Goldreich} \& {Sridhar}(1995)}]{Goldreich&Sridhar95}
{Goldreich}, P., \& {Sridhar}, S. 1995, \apj, 438, 763

\bibitem[{{Grasdalen}(1979)}]{Grasdalen79}
{Grasdalen}, G.~L. 1979, \pasp, 91, 436

\bibitem[{{Green} {et~al.}(2004){Green}, {Tuffs}, \& {Popescu}}]{Green+04}
{Green}, D.~A., {Tuffs}, R.~J., \& {Popescu}, C.~C. 2004, \mnras, 355, 1315

\bibitem[{{Hibschman} \& {Arons}(2001)}]{Hibschman&Arons01}
{Hibschman}, J.~A., \& {Arons}, J. 2001, \apj, 560, 871

\bibitem[{{Inoue} {et~al.}(2011){Inoue}, {Asano}, \& {Ioka}}]{Inoue+11}
{Inoue}, T., {Asano}, K., \& {Ioka}, K. 2011, \apj, 734, 77

\bibitem[{{Kakuwa} {et~al.}(2015){Kakuwa}, {Toma}, {Asano}, {Kusunose}, \&
  {Takahara}}]{Kakuwa+15}
{Kakuwa}, J., {Toma}, K., {Asano}, K., {Kusunose}, M., \& {Takahara}, F. 2015,
  \mnras, 449, 551

\bibitem[{{Kennel} \& {Coroniti}(1984{\natexlab{a}})}]{Kennel&Coroniti84a}
{Kennel}, C.~F., \& {Coroniti}, F.~V. 1984{\natexlab{a}}, \apj, 283, 694

\bibitem[{{Kennel} \& {Coroniti}(1984{\natexlab{b}})}]{Kennel&Coroniti84b}
---. 1984{\natexlab{b}}, \apj, 283, 710

\bibitem[{{Komissarov}(2013)}]{Komissarov13}
{Komissarov}, S.~S. 2013, \mnras, 428, 2459

\bibitem[{{Komissarov} \& {Lyubarsky}(2004)}]{Komissarov&Lyubarsky04}
{Komissarov}, S.~S., \& {Lyubarsky}, Y.~E. 2004, \mnras, 349, 779

\bibitem[{{Kuiper} {et~al.}(2001){Kuiper}, {Hermsen}, {Cusumano}, {Diehl},
  {Sch{\"o}nfelder}, {Strong}, {Bennett}, \& {McConnell}}]{Kuiper+01}
{Kuiper}, L., {Hermsen}, W., {Cusumano}, G., {et~al.} 2001, \aap, 378, 918

\bibitem[{{Lynn} {et~al.}(2014){Lynn}, {Quataert}, {Chandran}, \&
  {Parrish}}]{Lynn+14}
{Lynn}, J.~W., {Quataert}, E., {Chandran}, B.~D.~G., \& {Parrish}, I.~J. 2014,
  \apj, 791, 71

\bibitem[{{Lyutikov}(2003)}]{Lyutikov03}
{Lyutikov}, M. 2003, \mnras, 339, 623

\bibitem[{{Mac{\'{\i}}as-P{\'e}rez} {et~al.}(2010){Mac{\'{\i}}as-P{\'e}rez},
  {Mayet}, {Aumont}, \& {D{\'e}sert}}]{Macias-Perez+10}
{Mac{\'{\i}}as-P{\'e}rez}, J.~F., {Mayet}, F., {Aumont}, J., \& {D{\'e}sert},
  F.-X. 2010, \apj, 711, 417

\bibitem[{{Mart{\'{\i}}n} {et~al.}(2014){Mart{\'{\i}}n}, {Torres}, {Cillis}, \&
  {de O{\~n}a Wilhelmi}}]{Martin+14}
{Mart{\'{\i}}n}, J., {Torres}, D.~F., {Cillis}, A., \& {de O{\~n}a Wilhelmi},
  E. 2014, \mnras, 443, 138

\bibitem[{{Mart{\'{\i}}n} {et~al.}(2016){Mart{\'{\i}}n}, {Torres}, \&
  {Pedaletti}}]{Martin+16}
{Mart{\'{\i}}n}, J., {Torres}, D.~F., \& {Pedaletti}, G. 2016, \mnras, 459,
  3868

\bibitem[{{Mart{\'{\i}}n} {et~al.}(2012){Mart{\'{\i}}n}, {Torres}, \&
  {Rea}}]{Martin+12}
{Mart{\'{\i}}n}, J., {Torres}, D.~F., \& {Rea}, N. 2012, \mnras, 427, 415

\bibitem[{{Medin} \& {Lai}(2010)}]{Medin&Lai10}
{Medin}, Z., \& {Lai}, D. 2010, \mnras, 406, 1379

\bibitem[{{Mertsch} \& {Sarkar}(2011)}]{Mertsch&Sarkar11}
{Mertsch}, P., \& {Sarkar}, S. 2011, Physical Review Letters, 107, 091101

\bibitem[{{Metzger} {et~al.}(2014){Metzger}, {Vurm}, {Hasco{\"e}t}, \&
  {Beloborodov}}]{Metzger+14}
{Metzger}, B.~D., {Vurm}, I., {Hasco{\"e}t}, R., \& {Beloborodov}, A.~M. 2014,
  \mnras, 437, 703

\bibitem[{{Moran} {et~al.}(2013){Moran}, {Shearer}, {Mignani},
  {S{\l}owikowska}, {De Luca}, {Gouiff{\`e}s}, \& {Laurent}}]{Moran+13}
{Moran}, P., {Shearer}, A., {Mignani}, R.~P., {et~al.} 2013, \mnras, 433, 2564

\bibitem[{{Morlino} {et~al.}(2015){Morlino}, {Lyutikov}, \&
  {Vorster}}]{Morlino+15}
{Morlino}, G., {Lyutikov}, M., \& {Vorster}, M. 2015, \mnras, 454, 3886

\bibitem[{{Murase} {et~al.}(2015){Murase}, {Kashiyama}, {Kiuchi}, \&
  {Bartos}}]{Murase+15}
{Murase}, K., {Kashiyama}, K., {Kiuchi}, K., \& {Bartos}, I. 2015, \apj, 805,
  82

\bibitem[{{Ney} \& {Stein}(1968)}]{Ney&Stein68}
{Ney}, E.~P., \& {Stein}, W.~A. 1968, \apjl, 152, L21

\bibitem[{{Olmi} {et~al.}(2014){Olmi}, {Del Zanna}, {Amato}, {Bandiera}, \&
  {Bucciantini}}]{Olmi+14}
{Olmi}, B., {Del Zanna}, L., {Amato}, E., {Bandiera}, R., \& {Bucciantini}, N.
  2014, \mnras, 438, 1518

\bibitem[{{Owen} \& {Barlow}(2015)}]{Owen&Barlow15}
{Owen}, P.~J., \& {Barlow}, M.~J. 2015, \apj, 801, 141

\bibitem[{{Porth} {et~al.}(2014{\natexlab{a}}){Porth}, {Komissarov}, \&
  {Keppens}}]{Porth+14b}
{Porth}, O., {Komissarov}, S.~S., \& {Keppens}, R. 2014{\natexlab{a}}, \mnras,
  443, 547

\bibitem[{{Porth} {et~al.}(2014{\natexlab{b}}){Porth}, {Komissarov}, \&
  {Keppens}}]{Porth+14a}
---. 2014{\natexlab{b}}, \mnras, 438, 278

\bibitem[{{Porth} {et~al.}(2016){Porth}, {Vorster}, {Lyutikov}, \&
  {Engelbrecht}}]{Porth+16}
{Porth}, O., {Vorster}, M.~J., {Lyutikov}, M., \& {Engelbrecht}, N.~E. 2016,
  \mnras, 460, 4135

\bibitem[{{Ptuskin}(1988)}]{Ptuskin88}
{Ptuskin}, V.~S. 1988, Soviet Astronomy Letters, 14, 255

\bibitem[{{Rees} \& {Gunn}(1974)}]{Rees&Gunn74}
{Rees}, M.~J., \& {Gunn}, J.~E. 1974, \mnras, 167, 1

\bibitem[{{Reynolds}(2009)}]{Reynolds09}
{Reynolds}, S.~P. 2009, \apj, 703, 662

\bibitem[{{Sasaki} {et~al.}(2015){Sasaki}, {Asano}, \& {Terasawa}}]{Sasaki+15}
{Sasaki}, K., {Asano}, K., \& {Terasawa}, T. 2015, \apj, 814, 93

\bibitem[{{Schlickeiser}(1989)}]{Schlickeiser89}
{Schlickeiser}, R. 1989, \apj, 336, 243

\bibitem[{{Schlickeiser} \& {Miller}(1998)}]{Schlickeiser&Miller98}
{Schlickeiser}, R., \& {Miller}, J.~A. 1998, \apj, 492, 352

\bibitem[{{Shibata} {et~al.}(2003){Shibata}, {Tomatsuri}, {Shimanuki}, {Saito},
  \& {Mori}}]{Shibata+03}
{Shibata}, S., {Tomatsuri}, H., {Shimanuki}, M., {Saito}, K., \& {Mori}, K.
  2003, \mnras, 346, 841

\bibitem[{{Sironi} \& {Spitkovsky}(2011)}]{Sironi&Spitkovsky11}
{Sironi}, L., \& {Spitkovsky}, A. 2011, \apj, 741, 39

\bibitem[{{Slane} {et~al.}(2000){Slane}, {Chen}, {Schulz}, {Seward}, {Hughes},
  \& {Gaensler}}]{Slane+00}
{Slane}, P., {Chen}, Y., {Schulz}, N.~S., {et~al.} 2000, \apjl, 533, L29

\bibitem[{{Slane} {et~al.}(2004){Slane}, {Helfand}, {van der Swaluw}, \&
  {Murray}}]{Slane+04}
{Slane}, P., {Helfand}, D.~J., {van der Swaluw}, E., \& {Murray}, S.~S. 2004,
  \apj, 616, 403

\bibitem[{{Smith}(2003)}]{Smith03}
{Smith}, N. 2003, \mnras, 346, 885

\bibitem[{{Tanaka}(2013)}]{Tanaka13}
{Tanaka}, S.~J. 2013, in IAU Symposium, Vol. 291, Neutron Stars and Pulsars:
  Challenges and Opportunities after 80 years, ed. J.~{van Leeuwen}, 511--513

\bibitem[{{Tanaka}(2016)}]{Tanaka16}
{Tanaka}, S.~J. 2016, \apj, 827, 135

\bibitem[{{Tanaka} \& {Takahara}(2010)}]{Tanaka&Takahara10}
{Tanaka}, S.~J., \& {Takahara}, F. 2010, \apj, 715, 1248

\bibitem[{{Tanaka} \& {Takahara}(2011)}]{Tanaka&Takahara11}
---. 2011, \apj, 741, 40

\bibitem[{{Tanaka} \& {Takahara}(2013{\natexlab{a}})}]{Tanaka&Takahara13a}
---. 2013{\natexlab{a}}, Progress of Theoretical and Experimental Physics, 12,
  3

\bibitem[{{Tanaka} \& {Takahara}(2013{\natexlab{b}})}]{Tanaka&Takahara13b}
---. 2013{\natexlab{b}}, \mnras, 429, 2945

\bibitem[{{Tang} \& {Chevalier}(2012)}]{Tang&Chevalier12}
{Tang}, X., \& {Chevalier}, R.~A. 2012, \apj, 752, 83

\bibitem[{{Temim} {et~al.}(2012){Temim}, {Sonneborn}, {Dwek}, {Arendt},
  {Gehrz}, {Slane}, \& {Roellig}}]{Temim+12}
{Temim}, T., {Sonneborn}, G., {Dwek}, E., {et~al.} 2012, \apj, 753, 72

\bibitem[{{Temim} {et~al.}(2006){Temim}, {Gehrz}, {Woodward}, {Roellig},
  {Smith}, {Rudnick}, {Polomski}, {Davidson}, {Yuen}, \& {Onaka}}]{Temim+06}
{Temim}, T., {Gehrz}, R.~D., {Woodward}, C.~E., {et~al.} 2006, \aj, 132, 1610

\bibitem[{{Timokhin} \& {Harding}(2015)}]{Timokhin&Harding15}
{Timokhin}, A.~N., \& {Harding}, A.~K. 2015, \apj, 810, 144

\bibitem[{{Torres} {et~al.}(2014){Torres}, {Cillis}, {Mart{\'{\i}}n}, \& {de
  O{\~n}a Wilhelmi}}]{Torres+14}
{Torres}, D.~F., {Cillis}, A., {Mart{\'{\i}}n}, J., \& {de O{\~n}a Wilhelmi},
  E. 2014, Journal of High Energy Astrophysics, 1, 31

\bibitem[{{Torres} {et~al.}(2013{\natexlab{a}}){Torres}, {Cillis}, \&
  {Mart{\'{\i}}n Rodriguez}}]{Torres+13a}
{Torres}, D.~F., {Cillis}, A.~N., \& {Mart{\'{\i}}n Rodriguez}, J.
  2013{\natexlab{a}}, \apjl, 763, L4

\bibitem[{{Torres} {et~al.}(2013{\natexlab{b}}){Torres}, {Mart{\'{\i}}n}, {de
  O{\~n}a Wilhelmi}, \& {Cillis}}]{Torres+13b}
{Torres}, D.~F., {Mart{\'{\i}}n}, J., {de O{\~n}a Wilhelmi}, E., \& {Cillis},
  A. 2013{\natexlab{b}}, \mnras, 436, 3112

\bibitem[{{van der Swaluw} {et~al.}(2001){van der Swaluw}, {Achterberg},
  {Gallant}, \& {T{\'o}th}}]{vanderSwaluw+01}
{van der Swaluw}, E., {Achterberg}, A., {Gallant}, Y.~A., \& {T{\'o}th}, G.
  2001, \aap, 380, 309

\bibitem[{{Venter} \& {de Jager}(2007)}]{Venter&deJager07}
{Venter}, C., \& {de Jager}, O.~C. 2007, in WE-Heraeus Seminar on Neutron Stars
  and Pulsars 40 years after the Discovery, ed. W.~{Becker} \& H.~H. {Huang},
  40

\bibitem[{{Vinyaikin}(2007)}]{Vinyaikin07}
{Vinyaikin}, E.~N. 2007, Astronomy Reports, 51, 570

\bibitem[{{Volpi} {et~al.}(2008){Volpi}, {Del Zanna}, {Amato}, \&
  {Bucciantini}}]{Volpi+08}
{Volpi}, D., {Del Zanna}, L., {Amato}, E., \& {Bucciantini}, N. 2008, \aap,
  485, 337

\bibitem[{{Vorster} \& {Moraal}(2013)}]{Vorster&Moraal13}
{Vorster}, M.~J., \& {Moraal}, H. 2013, \apj, 765, 30

\bibitem[{{Vorster} \& {Moraal}(2014)}]{Vorster&Moraal14}
---. 2014, \apj, 788, 132

\bibitem[{{Vorster} {et~al.}(2013){Vorster}, {Tibolla}, {Ferreira}, \&
  {Kaufmann}}]{Vorster+13}
{Vorster}, M.~J., {Tibolla}, O., {Ferreira}, S.~E.~S., \& {Kaufmann}, S. 2013,
  \apj, 773, 139

\bibitem[{{Weiler} \& {Panagia}(1978)}]{Weiler&Panagia78}
{Weiler}, K.~W., \& {Panagia}, N. 1978, \aap, 70, 419

\bibitem[{{Wilson-Hodge} {et~al.}(2011){Wilson-Hodge}, {Cherry}, {Case},
  {Baumgartner}, {Beklen}, {Narayana Bhat}, {Briggs}, {Camero-Arranz},
  {Chaplin}, {Connaughton}, {Finger}, {Gehrels}, {Greiner}, {Jahoda}, {Jenke},
  {Kippen}, {Kouveliotou}, {Krimm}, {Kuulkers}, {Lund}, {Meegan}, {Natalucci},
  {Paciesas}, {Preece}, {Rodi}, {Shaposhnikov}, {Skinner}, {Swartz}, {von
  Kienlin}, {Diehl}, \& {Zhang}}]{Wilson-Hodge+11}
{Wilson-Hodge}, C.~A., {Cherry}, M.~L., {Case}, G.~L., {et~al.} 2011, \apjl,
  727, L40

\bibitem[{{Zhang} {et~al.}(2008){Zhang}, {Chen}, \& {Fang}}]{Zhang+08}
{Zhang}, L., {Chen}, S.~B., \& {Fang}, J. 2008, \apj, 676, 1210

\bibitem[{{Zhu} {et~al.}(2015){Zhu}, {Fang}, \& {Zhang}}]{Zhu+15}
{Zhu}, B.-T., {Fang}, J., \& {Zhang}, L. 2015, \mnras, 451, 3145

\end{thebibliography}

\end{document}